\documentclass[twocolumn]{aastex631}

\shorttitle{Abundances of Exoplanet Hosting M dwarfs: Future JWST Targets}
\shortauthors{Melo et al.}

\graphicspath{{./}{figures/}}

\begin{document}

\title{Stellar Characterization and Chemical Abundances of Exoplanet Hosting M dwarfs from APOGEE Spectra: Future JWST Targets}

\correspondingauthor{Edypo Melo}
\email{edyporibeiro@academico.ufs.br}

\author[0009-0007-9388-1634]{Edypo Melo}
\affiliation{Departamento de F\'isica, Universidade Federal de Sergipe, Av. Marcelo Deda Chagas, S/N Cep 49.107-230, S\~ao Crist\'ov\~ao, SE, Brazil}

\author[0000-0002-7883-5425]{Diogo Souto}
\affiliation{Departamento de F\'isica, Universidade Federal de Sergipe, Av. Marcelo Deda Chagas, S/N Cep 49.107-230, S\~ao Crist\'ov\~ao, SE, Brazil}

\author[0000-0001-6476-0576]{Katia Cunha}
\affiliation{Observatório Nacional/MCTIC, R. Gen. José Cristino, 77,  20921-400, Rio de Janeiro, Brazil}
\affiliation{Steward Observatory, University of Arizona, 933 North Cherry Avenue, Tucson, AZ 85721-0065, USA}

\author[0000-0002-0134-2024]{Verne V. Smith}
\affiliation{NSF’s NOIRLab, 950 N. Cherry Ave. Tucson, AZ 85719 USA}

\author[0000-0003-0697-2209]{Fábio Wanderley}
\affiliation{Observatório Nacional/MCTIC, R. Gen. José Cristino, 77, 20921-400, Rio de Janeiro, Brazil}

\author[0000-0002-9612-8054]{Vinicius Grilo}
\affiliation{Departamento de F\'isica, Universidade Federal de Sergipe, Av. Marcelo Deda Chagas, S/N Cep 49.107-230, S\~ao Crist\'ov\~ao, SE, Brazil}

\author[0009-0007-2704-6198]{Deusalete Camara}
\affiliation{Departamento de F\'isica, Universidade Federal de Sergipe, Av. Marcelo Deda Chagas, S/N Cep 49.107-230, S\~ao Crist\'ov\~ao, SE, Brazil}

\author[0009-0002-9223-9467]{Kely Murta}
\affiliation{Departamento de F\'isica, Universidade Federal de Sergipe, Av. Marcelo Deda Chagas, S/N Cep 49.107-230, S\~ao Crist\'ov\~ao, SE, Brazil}

\author[0000-0001-5541-6087]{Neda Hejazi}
\affiliation{Department of Physics and Astronomy, University of Kansas, Lawrence, KS, USA}

\author[0000-0002-1835-1891]{Ian J.M.\ Crossfield}
\affiliation{Department of Physics and Astronomy, University of Kansas, Lawrence, KS, USA}

\author[0009-0008-2801-5040]{Johanna Teske}
\affiliation{Earth \& Planets Laboratory, Carnegie Institution for Science, 5241 Broad Branch Road, NW, Washington, DC 20015}

\author[0000-0002-4671-2957]{Rafael Luque}
\affiliation{Department of Astronomy \& Astrophysics, University of Chicago, Chicago, IL, USA}

\author[0000-0002-0659-1783]{Michael Zhang}
\affiliation{Department of Astronomy \& Astrophysics, University of Chicago, Chicago, IL, USA}

\author[0000-0003-4733-6532]{Jacob Bean}
\affiliation{Department of Astronomy \& Astrophysics, University of Chicago, Chicago, IL, USA}

\begin{abstract}
% \textbf{\ds{An understanding of stellar atmospheric parameters and abundances is crucial to enhancing the interpretation of exoplanetary atmosphere retrievals.}} 
Exoplanets hosting M dwarfs are the best targets to characterize Earth-like or super-Earth planetary atmospheres with the James Webb Space Telescope (JWST). 
We determine detailed stellar parameters ($T_{\rm eff}$, log$g$, and $\xi$) and individual abundances of twelve elements for four cool M dwarfs hosting exoplanets TOI-1685, GJ 436, GJ 3470, and TOI-2445, scheduled for future observations by the JWST. The analysis utilizes high-resolution near-infrared spectra from the SDSS-IV APOGEE survey between 1.51-1.69$\micron$. Based on 1D-LTE plane-parallel models, we find that TOI-2445 is slightly metal-poor ([Fe/H] = -0.16$\pm$0.09), while TOI-1685, GJ 436 and GJ 3470 are more metal-rich ([Fe/H] = 0.06$\pm$0.18, 0.10$\pm$0.20 dex, 0.25$\pm$0.15). The derived C/O ratios for TOI-2445, TOI-1685, GJ 436, and GJ 3470 are 0.526$\pm$0.027, 0.558$\pm$0.097, 0.561$\pm$0.029, and 0.638$\pm$0.015, respectively. From results for 28 M dwarfs analyzed homogeneously from APOGEE spectra, we find exoplanet-hosting M dwarfs exhibit a C/O abundance ratio approximately 0.01 to 0.05 higher than those with non-detected exoplanets, at limits of a statistically significant offset. A linear regression of [Fe/H] \textit{vs.} C/O distribution reveals a noticeable difference in the angular coefficient between FGK dwarfs (0.27) and M dwarfs (0.13). Assuming our abundance ratios of Ca/Mg, Si/Mg, Al/Mg, and Fe/Mg, we determine a mass of 3.276$^{+0.448}_{-0.419}$$M_{\earth}$ for TOI-2445 b, having density (6.793$^{+0.005}_{-0.099}$ g.cm$^{-3}$) and core mass fraction (0.329$_{-0.049}^{+0.028}$) very similar to Earth's. We also present an atlas of 113 well-defined spectral lines to analyze M dwarfs in the $H$-band and a comprehensive evaluation of uncertainties from variations in the atmospheric parameters, signal-to-noise, and pseudo-continuum.
% \textbf{\ds{The work presented herein paves the way for future detailed studies of the chemistry of M dwarf small planet systems.}}
\end{abstract}


\section{Introduction} \label{sec:intro}

% Exoplanets orbiting M dwarf stars are prime candidates for characterizing Earth-like or \textbf{s}uper-Earth atmospheres using the James Webb Space Telescope (JWST). The choice of M dwarf targets reflects an 
% effort to enhance our understanding of the intricate relationship between small stars and smaller exoplanets. 
% This work focuses on characterizing exoplanet-hosting M dwarfs. 
M-dwarf stars, characterized by their low masses, small radii, and low effective temperatures ($T_{\rm eff}$), constitute a substantial portion of the Milky Way's stellar population, accounting for approximately 70\% of all stars (\citealt{Salpeter1955}; \citealt{Miller1979}; \citealt{Henry2018}). Despite their prevalence, M dwarfs have remained among the least-studied stellar types regarding their chemical abundances. The intricate optical spectra of M dwarfs, filled with strong molecular bands such as TiO and VO (\citealt{Allard2000}), have presented significant challenges to unveiling their chemical compositions. 
An increased interest in studying M dwarf chemical makeup has risen recently, motivated by the growing discovery of Earth-sized exoplanets orbiting these stars and high-resolution instruments operating in the near-infrared regime (NIR). The advantageous mass and size ratios between small planets and M dwarfs make Earth-sized-mass exoplanets particularly amenable to detection through radial velocity or transit methods (\citealt{Charbonneau2007}; \citealt{Gaidos2007}; \citealt{Shields2016}, and \citealt{Dressing2015}). 
This fact, coupled with the need for precise stellar characterization in the star-planet connection, emphasizes the need to improve the methodology for M-dwarf spectral analysis.
This is also relevant because exoplanetary atmospheres are key targets for observations by the James Webb Space Telescope (JWST) (\citealt{Lustig-Yaeger2023} and \citealt{May2023}).

% M dwarf stars offer an advantageous window for spectroscopic analysis in the near-infrared (NIR) domain, where their spectra are less affected by molecular blends than in the optical range. 
The near-infrared (NIR) domain offers an advantageous window for spectroscopic analysis of M dwarf stars, as their spectra are less affected by molecular blends than in the optical range.
Within the $H$-band, between 1.51 and 1.69 \micron, studies like \cite{Souto2017, Souto2018Ross, Souto2020} have used the Apache Point Observatory Galactic Evolution Experiment (APOGEE, \citealt{Majewski2017}) high-resolution spectrograph (R $\sim$ 22,500) and demonstrated the feasibility of determining key atmospheric parameters ($T_{\rm eff}$, log $g$, metallicity, and microturbulence) as well as individual abundances for up to fourteen elements. Additionally, investigations by \cite{Birky2020}, \cite{Sarmento2021} (see also \citealt{Antoniadis-Karnavas2020}), and others have successfully determined atmospheric parameters for numerous targets using APOGEE spectra, with results aligning well with photometric calibrations and literature findings.

The CARMENES survey, operating in both optical (0.52-0.96 micron) and in the J and H bands at high resolution (R $\sim$ 100,000), has contributed to exoplanetary research and M dwarf spectroscopic characterization, finding 33 new planets, confirming 26 planet candidates from transits (\citealt{Ribas2023}), and characterizing their host star in detail (\citealt {Reiners2018}; \citealt{Passegger2018}). Notably, High-Resolution Cross-Correlation Spectroscopy (HRCCS) techniques, as exemplified in works such as \cite{Nortmann2018, Salz2018, Alonso-Floriano2019, Sanchez-Lopez2019}, have enabled the determination of abundances in the exoplanetary atmospheres themselves.
Using another NIR spectrograph, SPIRou, (\citealt{Donati2020}), \cite{Allart2019, Pelletier2021} have probed the atmospheres of exoplanets like WASP-107b and $\tau$ Bo\"otis b (\citealt{Allart2019, Pelletier2021}). 
Furthermore, the studies by \cite{Cristofari2023} and Wanderley et al. (2024, submitted) determined magnetic fields by analyzing Zeeman splitting in well-defined M dwarf NIR lines.
Magnetic fields in M dwarf stars have the potential to generate stellar spots and inhibit convection in the stellar interior. These effects can lead to an expansion of the stellar radii, causing stellar inflation and reducing its effective temperature \citep{Chabrier2007}. The recent work by \citet{Wanderley2023} investigated radius inflation in M dwarfs belonging to the Hyades open cluster, revealing that, on average, M dwarfs in the Hyades exhibit larger radii than anticipated based on stellar isochrones by about 2\%.

This work aims to characterize the atmospheric parameters and chemical abundances of four M dwarf stars hosting exoplanets, TOI-1685, TOI-2445, GJ 436, and GJ 3470. The planets TOI-1685 b and TOI-2445 b were discovered by the Transiting Exoplanet Survey Satellite (TESS) in 2021 (\citealt{Bluhm2021}) and 2022 (\citealt{Giacalone2022}), respectively, while the planet GJ 436 b was discovered by \cite{Butler2004} at the W. M. Keck Observatory in 2004, and the planet GJ 3470 b was discovered by \cite{Bonfils2012} using the HARPS spectrograph, located at the La Silla Observatory, in 2012.
The detailed analysis and characterization of these planet hosts are of particular interest as their exoplanets are scheduled for observations by JWST.

When we study the chemical composition of a star, we can infer that their respective exoplanets shared the same composition during their formation (although planetary formation processes may significantly alter the exoplanet abundances). The abundance distribution of the cloud that gave birth to the stars and planets plays a critical role in the formation and evolution of planetary systems (\citealt{Bond2010,Dorn2015}).
For example, the ratio of carbon to oxygen (C/O) in a star plays a crucial role in the chemistry of protoplanetary disks where planets form (\citealt{Kowska2023}) (see also \citealt{Nissen2013}, \citealt{DelgadoMena2021}). A high C/O value can lead to the formation of planets rich in carbonaceous compounds, while a lower value may favor the formation of planets with water-rich atmospheres (\citealt{Madhusudhan2012, Teske2014}). 
The abundance of elements like magnesium, silicon, and iron in a star can influence the composition of an exoplanet's core and mantle (\citealt{Bond2010}, \citealt{DelgadoMena2010}, \citealt{Thiabaud2015}, \citealt{Dorn2017}, \citealt{Santos2017}, and \citealt{Unterborn2017}). 
This information is essential for determining whether a planet is predominantly rocky, gaseous, or has an intermediate composition (see Plotynokov \& Valencia 2020; Schultze et al. 2020 \& 2024). The distribution of these elements in the planet affects its internal structure, density, and, ultimately, its potential to host life as we know it on Earth.
 
This paper is organized as follows. In Section 2, we present the observations of the four stellar systems scheduled for observations with JWST. Section 3 discusses the details of our data modeling approach. Section 4 presents and discusses the results, including the C/O abundance ratios obtained for the stars. Section 5 provides a summary of our results. 
Finally, in the Appendix, we evaluate the sensitivity of abundance determinations obtained from $H$-band spectra to variations in stellar parameters, such as $T_{\rm eff}$, log $g$, $\xi$, and metallicity ([Fe/H]), as well as the impact that the signal-to-noise ratio (SNR) fluctuations and pseudo-continuum adjustments have on the results. 

\section{Observations} \label{sec:style}

We cross-matched a list of potential exoplanets scheduled for JWST observations\footnote{www.stsci.edu/~nnikolov/TrExoLiSTS/JWST/trexolists.html.} with the APOGEE DR17 dataset. This resulted in a total of seventeen stars, seven of which are M dwarfs. Three of these stars exhibit low signal-to-noise ratios (SNR $<$ 20) in DR17, and one presents issues with the reduction as processed by the DR17 APOGEE pipeline. However, its DR 16 \citep{Ahumada2020} spectrum was found to be of good quality. The remaining four targets, all with high SNR, are analyzed in this study: TOI-1685, GJ 436, TOI-2445, and GJ 3470.

We use observational data collected using the SDSS APOGEE North spectrograph (\citealt{Wilson2018}). APOGEE is a cryogenic, multi-fiber high-resolution (R = 22,500) instrument with 300 fibers in the near-infrared (NIR) spanning a spectral range from 1.51 to 1.69 $\micron$ ($H$-band) mounted on a 2.5-meter telescope. This spectrograph was commissioned during the third iteration of the Sloan Digital Sky Survey (SDSS-III) and has been operational at the Apache Point Observatory (APO) in New Mexico, USA, and continues to be used to gather data as part of the Milky Way Mapper Survey (MWM; \citealt{Kollmeier2017}).

\section{Spectral lines and Data modeling} \label{sec:modeling}

We conduct spectral synthesis to model the APOGEE spectra. Our modeling approach in this study employs 1D plane-parallel LTE MARCS model atmospheres \citep{Gustafsson2008}. For radiative transfer calculations, we use the turbospectrum code (\citealt[]{AlvarezPlez1998} and \citealt{Plez2012}), supplemented by the BACCHUS wrapper \cite[]{Masseron2016} in manual mode for computing the abundances.  We adopt a microturbulence velocity equal to 1 km.s$^{-1}$ as suggested by \cite{Souto2017}. Atomic and molecular transition data are from the APOGEE line list \cite[]{Smith2021}, which was used to compute the latest SDSS-IV data release (DR17; \citealt{Abdurrouf2022}) and, so far, is being utilized in the SDSS-V (\citealt{Kollmeier2017}).

In this study, we analyze the behavior of 113 spectral lines, focusing on mid to early M dwarfs with effective temperatures between 3200 to 4000K, surface gravity values between 4.5 and 5.5 dex, metallicities spanning from [Fe/H]= -0.16 to +0.25 dex, and microturbulent velocities ranging between 0.5 to 1.5 km.s$^{-1}$. While most spectral lines are suitable for precise abundance determinations in early-type M dwarfs at $\sim$ 3900K, lower effective temperatures result in blending with molecular lines, particularly with H$_{2}$O and FeH. 
In Figure \ref{fig:geral}, we illustrate the sensitivity of spectral lines to atmospheric parameters within a 70-angstrom window spanning from 15720 to 15790 \AA, reflecting the change of lines in both depth and shape due to the variation of these parameters. This particular spectral region is highly responsive to changes in effective temperature, surface gravity, and metallicity, primarily due to the presence of three Mg I lines, marked in Figure \ref{fig:geral}.

Although the M dwarf stars studied here cover a somewhat narrow range in effective temperature and metallicity (Section 4, Table 3), Table \ref{tab:lines} provides a comprehensive list of those 113 well-defined spectral lines suitable for precise individual abundance determinations of M dwarfs in the $H$-band and in particular in the APOGEE region. This table is split into different regimes of $T_{\rm eff}$; each of the sections in the table indicates the lines available for the respective $T_{\rm eff}$ regime.
%Notably, the spectral lines of Ca I, K I, Al I, Mg I, Na I, CO, and H$_{2}$O are measurable across the %entire range of atmospheric parameters \textbf{of the target stars} %considered in this study. 
We use these lines to determine the abundances and uncertainties in this work but note that such lines can also be useful for spectral analyses from other high-resolution spectrographs covering the $H$-band, such as CRIRES, CARMENES, and SPIRou, although other well-defined lines might appear as the spectral resolution increases. 

\section{Abundance Analysis} \label{sec:abundance}

We examine a set of fourteen elements encompassing transitions from both neutral atomic (Fe I, Na I, Mg I, Al I, Si I, K I, Ca I, Ti I, V I, Cr I, Mn I, Ni I) and molecular lines (CO, OH, H$_{2}$O, and FeH). 
We determined atmospheric parameters, $T_{\rm eff}$ and log $g$, by combining H$_{2}$O and OH lines, ensuring self-consistent oxygen abundance values. For details of this methodology, we refer to \cite{Souto2020}. 

Our abundance analysis involves a careful look at each line available (see Table \ref{tab:lines}) to ensure that only well-defined lines are adopted. From a visual inspection, we remove lines that are either blended, too small, or affected by spectra reduction issues. This is particularly relevant because as the stellar effective temperature decreases, the water vapor (H$_{2}$O) and iron hydride lines (FeH) become notably more prominent, often blending those well-defined neutral lines (\citealt{Allard2000,Tsuji2015}) in the $H$-band spectra of cooler M dwarfs.
In this study, the stellar metallicity, [Fe/H], is the average iron abundances derived from Fe I and FeH line indicators.

\section{Results and Discussion} \label{sec:cite}

The effective temperatures, surface gravity values, metallicities, and individual abundances were calculated using the spectral lines provided in Table \ref{tab:lines}, and results are summarized in Table \ref{tab:abundances} and 3.
The uncertainties presented in Table \ref{tab:abundances} were computed by propagating their respective errors in the atmospheric parameters, SNR, and pseudo-continuum change (assumed as 1\%) from Tables \ref{tab:sensitivity}, \ref{tab:noise}, and \ref{tab:displacement} via a quadratic sum. 

\begin{deluxetable*}{lr}
%\rotate
\label{tab:lines}
\tabletypesize{\tiny}
\tablecaption{M dwarfs well-defined spectral lines in the $H$-band.}
\tablewidth{0pt}
\tablehead{
\colhead{Elements} &
\colhead{Lines (in \AA{})}
}
\startdata
  For $T_{\rm eff}$ from 4000 K to 3200 K \\
  Ca & 16136.823, 16150.763, 16157.364 \\  
  K & 15163.067, 15168.376 \\
  Al & 16718.957, 16750.564, 16763.360 \\
  Mg & 15740.716, 15748.988, 15765.842 \\
  Na & 16373.853, 16388.858 \\
  C & 15977.7, 16184.9 \\
  H$_{2}$O (for $T_{\rm eff}$ $<$ 3800K) & 15256.8, 15258.3, 15258.4, 15259.2, 15259.4, 15270.6, 15315.7,  15317.3, 15317.5, 15353.6, 15360.5 \\ & 15447.6, 15455.8, 15461.5, 15503.6  \\
  \hline  
  For $T_{\rm eff}$ = 4000 K to 3700 K \\
  Ni & 15605.68, 15632.654, 16584.439, 16589.295, 16673.711, 16815.471, 16818.76 \\  
  Mn & 15159.0, 15217.8, 15262.4 \\  
  Cr & 15680.063 \\
  V & 15924.9 \\
  Ti & 15334.847, 15543.756, 15602.842, 15698.979, 15715.573, 16635.161 \\
  Si & 15888.410, 15960.063, 16094.787, 16680.770 \\
  OH & 15391.208, 15407.288, 15409.308, 15505.782, 15558.023, 15560.244, 15565.961, 15568.780, 15572.084, 16052.765 \\ & 16055.464, 16061.7, 16065.054, 16069.524, 16074.163, 16190.263, 16192.208, 16204.076, 16207.186, 16352.217 \\ & 16354.582, 16364.590, 16368.135, 16581.250, 16582.013, 16866.688, 16871.895, 16879.090, 16884.530, 16886.279 \\ & 16895.180, 16898.887 \\
  FeI & 15207.5, 15219.5, 15244.8, 15294.6, 15395, 15490.3, 15591.8, 15604, 15621.7, 15632 \\ & 15648.5, 15662, 15692.5, 15723.5 \\
  FeH & 15965, 16009.6, 16018.5, 16108.1, 16114, 16245.7, 16271.8, 16284.7, 16299.4, 16377.4 \\ & 16546.8, 16548.8, 16557.2, 16574.8, 16694.4, 16735.4, 16738.3, 16741.7, 16796.4, 16812.7 \\ & 16814.1, 16889.6, 16892.9, 16922.7, 16935.1 \\
  \hline  
  For $T_{\rm eff}$ = 3600 K \\
% Ni &  \\  
 Mn & 15159 \\  
% Cr &  \\
 V & 15924.9 \\
 Ti & 15334.847, 15543.756, 15715.573 \\
 Si & 15888.410 \\
 OH & 15391.208, 15407.288, 15409.308, 15505.782, 15558.023, 15560.244, 15565.961, 15568.780, 15572.084, 16052.765 \\ & 16055.464, 16061.7, 16065.054, 16069.524, 16074.163, 16190.263, 16192.208, 16204.076, 16207.186, 16352.217 \\ & 16354.582, 16364.590, 16368.135, 16581.250, 16582.013, 16866.688, 16871.895, 16879.090, 16884.530, 16886.279 \\ & 16895.180, 16898.887 \\
 FeI & 15207.5, 15219.5, 15294.6, 15621.7, 15632, 15723.5 \\
 FeH & 15965, 16009.6, 16018.5, 16108.1, 16114, 16245.7, 16271.8, 16284.7, 16299.4, 16377.4 \\ & 16546.8, 16548.8, 16557.2, 16574.8, 16889.6, 16892.9, 16935.1 \\    
    \hline
  For $T_{\rm eff}$ = 3500 K \\
% Ni &  \\  
 Mn & 15159 \\  
% Cr &  \\
 V & 15924.9 \\
 Ti & 15334.847, 15543.756, 15715.573 \\
 Si & 15888.410 \\
 OH & 15391.208, 15407.288, 15409.308, 15505.782, 15558.023, 15560.244, 15565.961, 15568.780, 15572.084, 16052.765 \\ & 16055.464, 16061.7, 16065.054, 16069.524, 16074.163, 16190.263, 16192.208, 16204.076, 16207.186, 16352.217 \\ & 16354.582, 16364.590, 16368.135, 16581.250, 16582.013, 16866.688, 16871.895, 16879.090, 16884.530, 16886.279 \\ & 16895.180, 16898.887 \\
 FeI & 15294.6 \\
 FeH & 15965, 16009.6, 16018.5, 16108.1, 16114, 16245.7, 16271.8, 16284.7, 16299.4, 16377.4 \\ & 16546.8, 16548.8, 16557.2, 16574.8, 16889.6, 16892.9, 16935.1 \\    
  \hline
  For $T_{\rm eff}$ = 3400 K \\
% Ni &  \\  
% Mn &  \\  
% Cr &  \\
% V &  \\
 Ti & 15334.847, 15543.756, 15715.573 \\
 Si & 15888.410 \\
 OH & 15391.208, 15407.288, 15409.308, 15505.782, 15558.023, 15560.244, 15565.961, 15568.780, 15572.084, 16052.765 \\ & 16055.464, 16061.7, 16065.054, 16069.524, 16074.163, 16190.263, 16192.208, 16204.076, 16207.186, 16352.217 \\ & 16354.582, 16364.590, 16368.135, 16581.250, 16582.013, 16866.688, 16871.895, 16879.090, 16884.530, 16886.279 \\ & 16895.180, 16898.887 \\
 FeI & 15294.6 \\
 FeH & 15965, 16009.6, 16018.5, 16108.1, 16114, 16245.7, 16271.8, 16284.7, 16299.4, 16377.4 \\ & 16546.8, 16548.8, 16557.2, 16574.8, 16889.6, 16892.9, 16935.1 \\      
  \hline
  For $T_{\rm eff}$ = 3300 K \\
% Ni &  \\  
% Mn &  \\  
% Cr &  \\
% V &  \\
 Ti & 15334.847, 15543.756, 15715.573 \\
 Si & 15888.410 \\
 OH & 15391.208, 15407.288, 15409.308, 15505.782, 15558.023, 15560.244, 15565.961, 15568.780, 15572.084, 16052.765 \\ & 16055.464, 16061.7, 16065.054, 16069.524, 16074.163, 16190.263, 16192.208, 16204.076, 16207.186, 16352.217 \\ & 16354.582, 16364.590, 16368.135, 16581.250, 16582.013, 16866.688, 16871.895, 16879.090, 16884.530, 16886.279 \\ & 16895.180, 16898.887 \\
 FeI & 15294.6 \\
 FeH & 15965, 16009.6, 16018.5, 16108.1, 16114, 16245.7, 16271.8, 16284.7, 16299.4, 16377.4 \\ & 16546.8, 16548.8, 16557.2, 16574.8, 16889.6, 16892.9, 16935.1 \\      
  \hline
  For $T_{\rm eff}$ = 3200 K \\
%   Ni &  \\  
% Mn &  \\  
% Cr &  \\
% V &  \\
 Ti & 15334.847, 15543.756, 15715.573 \\
 Si & 15888.410 \\
 OH & 15391.208, 15407.288, 15409.308, 15505.782, 15558.023, 15560.244, 15565.961, 15568.780, 15572.084, 16052.765 \\ & 16055.464, 16061.7, 16065.054, 16069.524, 16074.163, 16190.263, 16192.208, 16204.076, 16207.186, 16866.688 \\ & 16871.895, 16879.090, 16884.530, 16886.279 \\
 FeI & 15294.6 \\
 FeH & 15965, 16009.6, 16018.5, 16108.1, 16114, 16245.7, 16271.8, 16284.7, 16299.4, 16377.4 \\ & 16546.8, 16548.8, 16557.2, 16574.8, 16889.6, 16892.9, 16935.1 \\ 
\tablewidth{0pt}	
\enddata
\end{deluxetable*}

%--------------------------------------------
\begin{figure*}
  \centering
  {\includegraphics[width=1.0\textwidth]{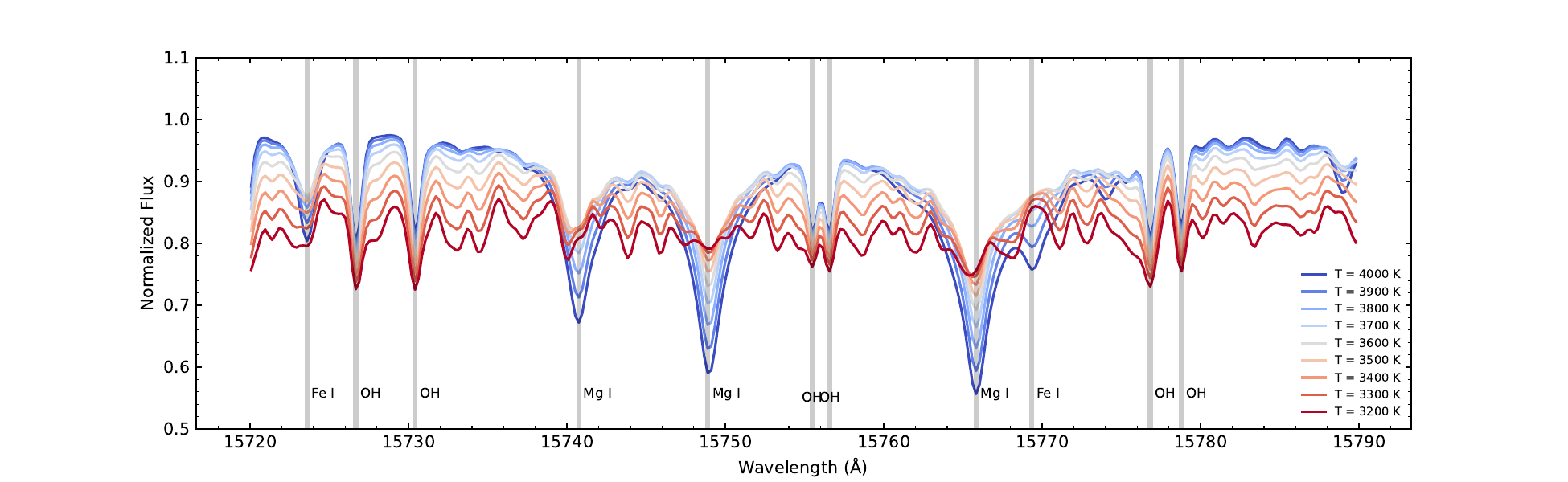}}\label{fig:a} \\
  {\includegraphics[width=1.0\textwidth]{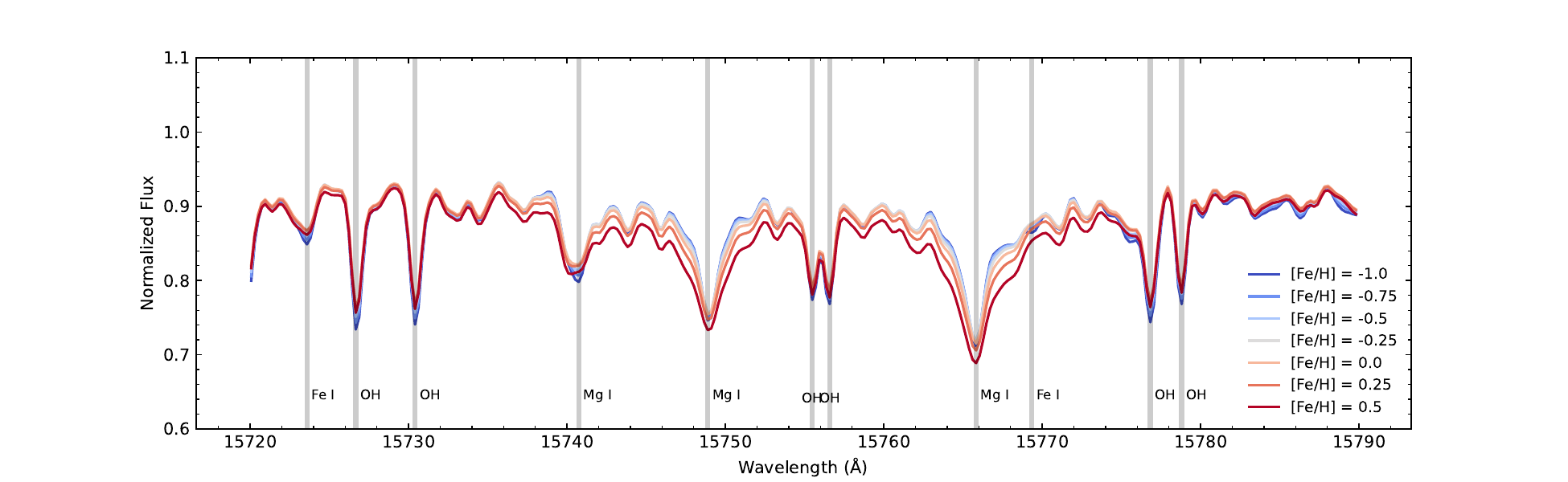}}\label{fig:b} \\
  {\includegraphics[width=1.0\textwidth]{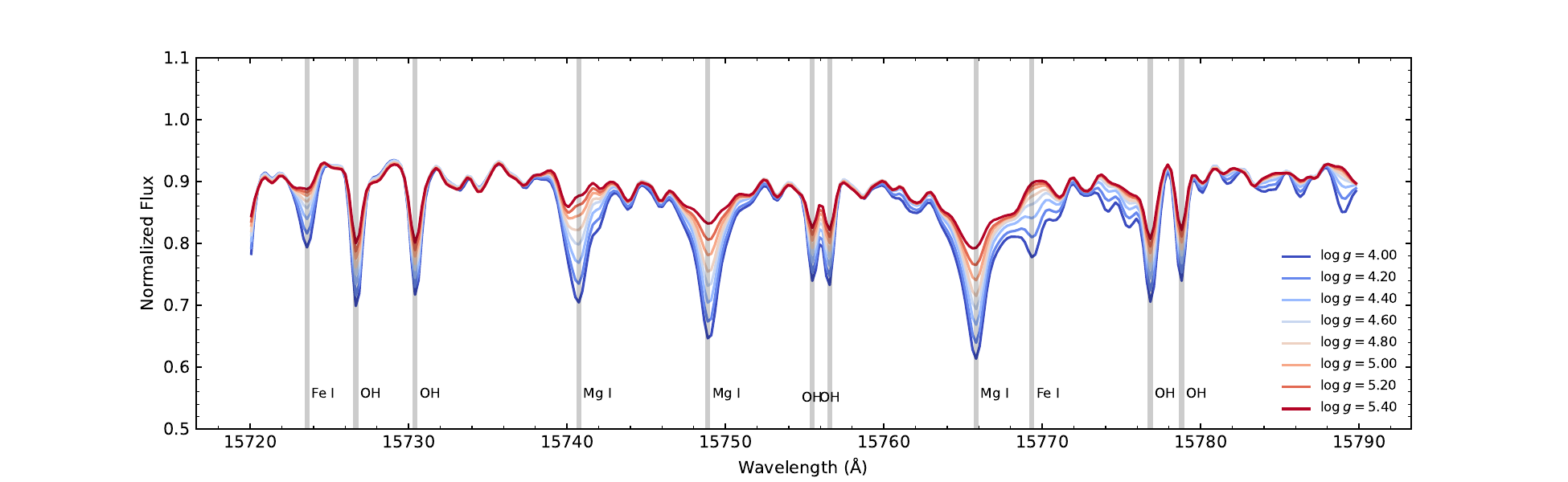}\label{fig:c}}
  \caption{Sensitivity to variations in stellar atmospheric parameters. 
  Top panel: spectral synthesis varying $T_{\rm eff}$ from 3200K (redder line) to 4000K (bluer line) in steps of 100K with fixed values of log $g$ (4.8 dex), [Fe/H] (0.00 dex), and $\xi$ (1.00 km.s$^{-1}$).
  Middle panel: spectral synthesis varying [Fe/H] from 0.5 (redder line) to -1.0 (bluer line) in steps of 0.25 dex with fixed values of $T_{\rm eff}$ (3500K), log $g$ (4.8 dex), and $\xi$ (1.00 km.s$^{-1}$).
  Bottom panel: spectral synthesis varying log $g$ from 5.4 (redder line) to 4.0 (bluer line) in steps of 0.2 dex with fixed values of $T_{\rm eff}$ (3500K), [Fe/H] (0.00 dex), and $\xi$ (1.00 km.s$^{-1}$).}
  \label{fig:geral}
\end{figure*}
%--------------------------------------------
%--------------------------------------------

\subsection{Exoplanet Hosting M dwarfs on JWST targeting list}

The search for and characterization of exoplanets are among the most exciting scientific endeavors in modern astronomy. To know an exoplanet well, we need to know its host star well. This is because the chemical composition of stars may play a fundamental role in understanding their exoplanets. Also, knowing the stellar mass, radius, and flux is critical because exoplanet physical properties rely on these parameters (such as M$_{planet}$/M$_{star}$, R$_{planet}$/R$_{star}$, F$_{planet}$/F$_{star}$). Furthermore, the connection between the stars and the planet's chemistry may enable understanding the nature and evolution of extraterrestrial worlds (\citealt{Kolecki2022,Polanski2022,Hejazi2023}). 

Based on the atmospheric parameters determined in this work, we can use physical relations or calibrations to obtain fundamental stellar properties, such as luminosity, radius, and mass. We determined the absolute magnitudes of the M-dwarfs TOI-1685, GJ 436, TOI-2445, and GJ 3470 using their respective apparent magnitudes from the K band (2MASS; \citealt{Cutri2002}) and distances from \cite{Gaia2022}, \cite{BailerJones2021}, \cite{Stassun2019}, \cite{Salz2015}, and \cite{Maciejewski2014}. 
We used the bolometric correction from \cite{Mann2015} to obtain the bolometric magnitude and determine the luminosity from physical relation. We adopt values from \cite{Mamajek2015} of M$_{bol}$ = 0.00, corresponding to L$_{0}$ = 3.0128.10$^{35}$ erg s$^{-1}$, which leads to a solar luminosity of 3.828.10$^{33}$ erg s$^{-1}$ and M$_{bol}$ (Sun) = 4.74. 
We determined the stellar radii and masses using calibrations from  \cite{Souto2020}, 
\begin{equation}
R_{\star}/R_{\odot} = \sum_{n=0}^{n} a_{n} \left({M_{K_{S}}} \right)^{n},
\end{equation}
where $a_{0} = 1.9932$, $a_{1} = -0.3659$, and $a_{2} = 0.0177$, and $M_{K_{S}}$ is the absolute $K_{S}$ magnitude. Stellar masses were computed from the equation
\begin{equation}
\frac{M_\star}{M_{\odot}} = 0.2524 - 0.5765 \left( \frac{R_{\star}}{R_{\odot}} \right) + 2.0122 \left( \frac{R_{\star}}{R_{\odot}} \right)^{2}.
\end{equation}
%Our 
The adopted photometry and derived stellar masses and radii are presented in Table \ref{tab:parameters}. The internal uncertainty in the derived stellar radii is 0.03 $R_{\star}$/$R_{\sun}$, while for the derived stellar masses, the internal uncertainty varies from 5 to 10$\%$ (about 0.02 to 0.05 $M_{\star}$/$M_{\sun}$).

We re-derived the planetary parameters with the stellar parameters derived in this study. We derived the planetary radii of TOI-1685 b, GJ 436 b, TOI-2445 b, and GJ 3470 b from the physical transit equation below, 

\begin{equation}
\delta = 1.049\left( \frac{R_p/R_{Jup}}{R_\star/R_\odot}\right)^2,
\end{equation}
with transit depths ($\delta$) provided by the Exoplanet Follow-up Observation Program - Transiting Exoplanet Survey Satellite \citep{Ricker2015} and \citep{Giacalone2022}.

For the determination of planetary masses ($M_p$), we used the radial velocity classical equation (see \citealt{Torres2008}), 
\begin{equation}
M_p \sin i = 4.919 \times 10^{-3} P^{1/3} (1 - e^2)^{1/2} K_\star \left( \frac{{M_\star + M_p}}{{M_\odot}} \right)^{2/3},
\end{equation}
where input parameters are our stellar masses, the planet orbital period ($P$), the radial velocity semi-amplitude ($K$; \cite{Kosiarek2019}, \cite{Hirano2021}, \cite{Rosenthal2021}), and the eccentricity (e; \cite{Kosiarek2019}), \cite{Rosenthal2021}. We assumed the inclination angle ($i$) is 90º. 

We also determined the exoplanet equilibrium temperature ($T_{eq}$) from the Stefan-Boltzman physical relation below (see \citealt{Lissauer2013} for details)
\begin{equation}
T_{eq} = T_{\text{eff}}\sqrt{\frac{R_{\star}}{2a}}(1 - A_{B})^{\frac{1}{4}},
\end{equation} 
assuming an approximation of no atmosphere and an albedo of 0.306. Table \ref{tab:parameters} summarizes the exoplanetary parameters adopted in the calculations and those derived in this study. In the following section, we discuss the results for each star-planet system.

\begin{deluxetable*}{lrrrr}
%\rotate
% \tablenum{5}
\label{tab:abundances}
\tabletypesize{\normalsize}
\tablecaption{Abundances of exoplanet Hosting M dwarfs on JWST targeting list.}
\tablewidth{0pt}
\tablehead{
\colhead{Element} &
\colhead{TOI-1685} &
\colhead{GJ 436} &
\colhead{TOI-2445} &
\colhead{GJ 3470}
}
\startdata
$[$Fe/H$]$ & 0.06$\pm$0.178 & 0.10$\pm$0.199  & -0.16$\pm$0.094 & 0.25$\pm$0.152 \\
$[$C/Fe$]$ & -0.15$\pm$0.108 & -0.07$\pm$0.146  & 0.08$\pm$0.110 & -0.05$\pm$0.104 \\
$[$O/Fe$]$ & -0.12$\pm$0.093 & -0.05$\pm$0.093  & 0.13$\pm$0.098 & -0.08$\pm$0.097\\
$[$Na/Fe$]$ & 0.03$\pm$0.118 & -- & -- & -0.03$\pm$0.120\\
$[$Mg/Fe$]$ & -0.05$\pm$0.172 & 0.00$\pm$0.178  & 0.01$\pm$0.144 & -0.07$\pm$0.159 \\
$[$Al/Fe$]$ & -0.10$\pm$0.095 & 0.00$\pm$0.100  & -0.05$\pm$0.084 & -0.16$\pm$0.097 \\
$[$Si/Fe$]$ & -0.11$\pm$0.235 & 0.01$\pm$0.264  & -- & -0.05$\pm$0.206\\
$[$K/Fe$]$ & -0.03$\pm$0.063 & -0.11$\pm$0.075  & -0.04$\pm$0.083 & -0.16$\pm$0.057 \\
$[$Ca/Fe$]$ & 0.04$\pm$0.083 & -0.03$\pm$0.090  & 0.06$\pm$0.121 & 0.00$\pm$0.074 \\
$[$Ti/Fe$]$ & -0.26$\pm$0.084 & -0.20$\pm$0.089  & -0.25$\pm$0.101 & -0.26$\pm$0.088 \\
$[$V/Fe$]$ & -0.18$\pm$0.250 & --& -- & -0.09$\pm$0.183 \\
$[$Mn/Fe$]$ & 0.00$\pm$0.040 & 0.04$\pm$0.167  & -- & -0.08$\pm$0.133\\
C/O         & 0.558$\pm$0.027	 & 0.561$\pm$0.097 & 0.526$\pm$0.029 &  0.638$\pm$0.015 \\
\tablewidth{0pt}	
\enddata
\end{deluxetable*}

\begin{deluxetable*}{lrrrr}
%\rotate
% \tablenum{5}
\label{tab:parameters}
\tabletypesize{\small}
\tablecaption{Stellar and exoplanetary parameters.}
\tablewidth{0pt}
\tablehead{
\colhead{} &
\colhead{TOI-1685 (b)} &
\colhead{GJ 436 (b)} &
\colhead{TOI-2445 (b)} &
\colhead{GJ 3470 (b)}
}
\startdata
Stellar parameters \\
M$_k$  & 5.882 & 6.127 & 7.347 & 5.646 \\
BC$_k$ & 2.689 & 2.694 & 2.718 & 2.673 \\
M$_{bol}$ & 8.571 & 8.821 & 10.065 & 8.318\\
$L$/$L$$_{\odot}$ & 0.029 & 0.023 & 0.007 & 0.037 \\
$R$/$R$$_{\odot}$ & 0.453$\pm$0.014 & 0.416$\pm$0.013 & 0.260$\pm$0.08 & 0.492$\pm$0.015 \\
$M$/$M$$_{\odot}$ & 0.405$\pm$0.041 & 0.361$\pm$0.036 & 0.239$\pm$0.024 & 0.455$\pm$0.046 \\
$T_{\rm eff}$ ($K$) & 3519$\pm$100 & 3507$\pm$100 & 3318$\pm$100 & 3640$\pm$100 \\
log $g$ & 4.63$\pm$0.13 & 4.81$\pm$0.13 & 4.94$\pm$0.13 & 4.79$\pm$0.13 \\
distance (pc) & 37.60 & 9.75 & 48.58 & 29.42 \\
V & 13.378 & 10.670 & 15.692 & 12.332 \\
J & 9.616 & 6.900 & 11.555 & 8.794 \\
H & 9.005 & 6.319 & 11.033 & 8.206 \\
K & 8.758 & 6.073 & 10.779 & 7.989 \\
Exoplanetary parameters \\
Transit Depth ($\delta$ $\%$) & 0.102 & 0.691 & 0.182 & 0.569 \\
Semi Major axis ($a$ au) & 0.012 & 0.029 & 0.006 & 0.036 \\
Period ($P$ days) & 0.669 & 2.644 & 0.371 & 3.337 \\
Radial velocity amplitude ($K$ m.s$^{-1}$) & 4.200 & 17.120 & -- & 8.210 \\
Eccentricity ($e$) & -- & 0.145 & -- & 0.114 \\
$R$/$R$$_{\earth}$ & 1.759$\pm$0.010 & 3.645$\pm$0.018 & 1.375$\pm$0.005 & 4.352$\pm$0.028 \\
$M$/$M$$_{\earth}$ & 3.142$\pm$0.807 & 18.548$\pm$0.405 & 3.276$\pm$0.434$^{\dagger}$ & 11.276$\pm$0.497 \\
$T_{\rm eq}$ ($K$) & 979$\pm$28 & 583$\pm$17 & 936$\pm$28 & 592$\pm$16 \\
\tablewidth{0pt}	
\enddata
\tablenotetext{\dagger}{Mass computed with ExoPlex \citep{Unterborn2017}.}
\end{deluxetable*}

% $^{+0.448}_{-0.419}$

\subsubsection{TOI-1685}

TOI-1685 (2MASS J04342248+4302148, TIC 28900646) is an M dwarf star located 37.6 parsecs away from our Sun (\citealt{Gaia2022}), having a spectral type M3.0 V, and belonging to the thin disk of our Galaxy (\citealt{Bluhm2021}). We determine TOI-1685 to have a $T_{\rm eff}$ of 3519$\pm$100K and log $g$ of 4.63$\pm$0.13 dex, and our effective temperature is slightly hotter but in agreement, within the uncertainties, with the results from \cite{Bluhm2021} of $T_{\rm eff}$= 3434$\pm51K$ and log $g$= 4.85$\pm0.04$ dex. 
The latter study measured stellar parameters and metallicities from the stacked CARMENES VIS spectra and from fits to a synthetic grid of PHOENIX-SESAM models (\citealt{Husser2013}, \citealt{Passegger2019}).
Our derived radius and mass are 0.453$\pm0.014 R_{\odot}$ and 0.405$\pm0.041 M_{\odot}$, respectively, while  the previous work by \citet{Bluhm2021} derived a mass of 0.495$\pm0.019 M_{\odot}$, and a radius of 0.492$\pm 0.015 R_{\odot}$ for TOI-1685. 

TOI-1685  exhibits a moderate rotation rate for being an early M dwarf.
Our best fit synthetic spectrum for this star was obtained for a rotational velocity (vsin $i$) of 7.0 km.s$^{-1}$, which is close to the threshold v sin $i$ that can be measured using APOGEE spectra. Concerning stellar activity, this star exhibits a pseudo-equivalent width, pEW(H$\alpha$), of $+0.51$ Å, which is indicative of it being active (\citealt{Cincunegui2007}),
but it exhibits a lack of X-ray and ultraviolet emission. Its estimated age is in the range between 0.6 and 2 Gyr,  based on gyrochronology relations and comparisons with open clusters like Praesepe (\citealt{Kraus2007}). 

We derived a near-solar metallicity for TOI-1685 of [Fe/H] = 0.06$\pm$0.18 dex. The previously mentioned study by \cite{Bluhm2021} found this star to be more metal-poor [Fe/H] = -0.13$\pm$0.16, while \cite{Hirano2021} reported a higher metallicity of [Fe/H] = 0.14$\pm$0.12 dex 
from a weighted average of the TRES spectra analysis by specMatch-Emp (\citealt{Yee2017}), the InfraRed Doppler (IRD) spectra (\citealt{Ishikawa2020}), and SED fitting with the NextGen model.
% from an analysis of TRES spectra, using specMatch-Emp code (\citealt{Yee2017}). 
Besides being observed by the APOGEE survey, TOI-1685 was also observed by the LAMOST survey, with reported metallicities in APOGEE DR17 of -0.05 (\citealt{Abdurrouf2022}) and LAMOST DR2 of -0.20 dex (\citealt{Ding2022}). 
It has been previously shown from APOGEE studies of M dwarfs in the Coma Berenice \citep{Souto2021} and Hyades \citep{Wanderley2023} open clusters that the metallicities obtained by the APOGEE pipeline for cooler M dwarfs in DR 17 are systematically lower than expected. Here, for TOI-1685, we find a $\sim$0.1 dex difference with the metallicity in DR17, a similar result within the uncertainties.
The discrepancy with the metallicity from LAMOST possibly highlights the challenges in accurately determining metallicities for low-mass stars at low resolution (R $\sim$ 1900). 

\cite{Hirano2021} work also reported abundances for seven elements, including Na, Mg, Si, Ca, Ti, and Mn. We find good agreement in Na and Ca abundances ($\delta$ this work - \citealt{Hirano2021}) $<$ 0.10 dex). However, our results for Mg, Si, and Mn show a systematic difference, with \cite{Hirano2021} reporting higher values, up to 0.50 dex. The largest difference is observed for Ti (0.83 dex), where \cite{Hirano2021} results appear to be notably higher.

% \cite{Hirano2021} also reported abundances for seven other elements. The ones in common with this work are Na, Mg, Si, Ca, Ti, and Mn. For Na and Ca, our results are in good agreement with \cite{Hirano2021} work ($\delta$ this work - \cite{Hirano2021}) $<$ 0.10 dex. However, for Mg, Si, and Mn, \cite{Hirano2021} results are systematically higher, up to 0.50 dex. An even higher difference is found for Ti (0.83 dex), but \cite{Hirano2021} results seem to be too high ([Ti/Fe] = 0.57).

TOI-1685 hosts at least one planet: TOI-1685 b, whose transits, eclipses, and thermal phase curve are being observed by JWST in GO programs 3263, 4098, and 4195 (\citealt{Luque2023,Benneke2023,Fisher2023}). It was discovered in 2021 using the transit method with the TESS satellite (\citealt{Ricker2015}). This planet is a ``super-Earth'' orbiting TOI-1685 at a close distance, just $0.012$ astronomical units (AU) away from the star (\citealt{Hirano2021}), and it completes one orbital period in 0.669 days (\citealt{Luque2022}).  
Our results for the mass and radius of TOI-1685 b are of 3.142$\pm 0.81M_{\earth}$ and 1.759$\pm 0.01R_\earth$, respectively, and this compares well with previous results from the literature (\citealt{Luque2022} obtained 
%With 
a mass of 3.09$\pm 0.58 M_\earth$ and a radius of about $1.70R_\earth$). TOI-1685 b boasts a high density of $4.122$ g.cm$^{-3}$, about $75\%$ of Earth's density (\citealt{Luque2022}). 
Compared to our Solar System, TOI-1685 b is significantly more massive than the Earth, larger in radius, and denser than Mars. Its proximity to its star resembles Mercury's distance from the Sun.

With the semi-major axis of the orbit provided in \cite{Hirano2021} and the effective temperature and stellar radius from this study, we obtain an equilibrium temperature for TOI-1685 b of 979$\pm 28 K$. 
The literature results for the equilibrium temperature for this planet are overall slightly higher than ours. The equilibrium temperature obtained by \cite{Bluhm2021} is 1069$\pm 16K$, and that obtained by \cite{Hirano2021} is 1052$\pm 26K$. 
We note, however, that our determination does not account for any atmosphere or greenhouse effect and that we assume an Earth-like albedo. 

\subsubsection{GJ 436}

Gliese 436 (GJ 436; J11421096+2642251) is a $M2.5V$ star situated 9.75 parsecs away from the Sun (\citealt{Salz2015}, \citealt{Maciejewski2014}).  GJ 436 has a mass of M=0.47$M_{\odot}$ and a radius of R=0.45$R_{\odot}$ (\citealt{Maciejewski2014}). 
\cite{Rosenthal2021} derives an effective temperature of $3586 K$, while \cite{Bourrier2018} report $T_{\rm eff}$ of 3479$\pm 60K$. 
Our determination for $T_{\rm eff}$ falls between these two studies, with $T_{\rm eff}$=3507$\pm 100 K$, while we derive log $g$ of 4.81$\pm 0.13$dex.  
Our results for the stellar radius and mass are R=0.417$\pm$ 0.013$R_{\odot}$ and M=0.362$\pm 0.036$$M_{\odot}$, respectively. Our analysis reveals that GJ 436 is slightly metal-rich, with [Fe/H] = +0.10$\pm$0.20. 
Using CARMENES spectra, \cite{Schweitzer2019} find GJ 436 metallicity to be [Fe/H]= -0.04$\pm$0.16, which is slightly more metal-poor than ours.
% The metallicity for this star obtained, for example, in the CARMENES study of \cite{Schweitzer2019} is more metal-poor than ours [Fe/H]= -0.04 $\pm$ 0.16. The results overlap with the uncertainties.

Comparing our findings with those of \cite{Ishikawa2022}, who also determined abundances for Fe, Mg, Si, K, Ca, Ti, and Mn using IRD spectra, we find good overall agreement within the uncertainties. The difference in [Fe/H] between our work and theirs is 0.02 dex. For Si, K, Ca, and Mn, the difference is less than 0.07 dex, indicating close agreement. However, we observe larger discrepancies for Mg and Ti, with differences of -0.18 and -0.41 dex, respectively. 
Additionally, our results align with those of \cite{Maldonado2020}, who utilized high-resolution optical spectra from HARPS-N to determine abundances for seven elements overlapping with our study: Fe, C, Mg, Al, Si, Ca, Ti, and Mn. The difference in [Fe/H] between our study and theirs is -0.09 dex. Notably, the difference for Al is -0.03 dex, while for Si, it is 0.01 dex. However, variations are more pronounced for C, Ti, Mn, Mg, and Ca, with differences of -0.11 dex, -0.13 dex, 0.09 dex, 0.21 dex, and 0.29 dex, respectively.

GJ 436 hosts at least one exoplanet, GJ 436 b, which was observed in secondary eclipse by the JWST GTO programs 1177 and 1185 (\citealt{Greene2017a}, and \citealt{Greene2017b}).  GJ 436 b is the prototypical exo-Neptune, as it was the first planet discovered with a size and mass comparable to Neptune: R=4.17$\pm$0.17$R_{\earth}$, with a mass that is approximately $22.1$ times that of Earth (\citealt{Butler2004}, \citealt{Gillon2007}).
The orbital semi-major axis of GJ 436 b is a=0.029 AU, with a period of P=2.644 days (\citealt{Maciejewski2014}). Our determinations for the planetary radius and mass are R=3.645$\pm$ 0.018$R_{\earth}$ and M=18.606$M_{\earth}$.

The planet has been subjected to atmospheric characterization via transit, eclipse, and thermal phase measurements from numerous observatories (\citealt{Stevenson2010}).  These results suggest that the planetary atmosphere is substantially deficient in methane relative to expectations from equilibrium chemistry and a solar-like, hydrogen-rich composition at the planet's expected temperature; the atmosphere is likely either significantly metal-rich ($\gtrsim$300$\times$ solar) and/or exhibits substantial disequilibrium chemistry (\citealt{Stevenson2010,Moses2013,Morley2017}).  Using the orbital semi-major axis from \cite{Maciejewski2014}, along with our effective temperature and stellar radius, we obtain an equilibrium temperature of GJ 436 b of 583$\pm 17 K$, while \cite{Turner2016} report a somewhat higher value of 686$\pm 10K$.
 
The mystery deepens as the planet is too compact to be primarily composed of hydrogen gas, similar to gas giants like Jupiter. Yet, it is not compact enough to be classified as a rocky super-Earth, having a density of 1.80$\pm 0.29$ g.cm$^{-3}$ (\citealt{Maciejewski2014}). One of the hypotheses regarding the composition of GJ 436 b suggests that it might predominantly consist of an exotic form of solid water. This unique water phase is believed to have solidified due to extreme pressure rather than low temperatures (\citealt{Gillon2007}). This well-known degeneracy in interpreting masses and radii of exo-Neptunes makes it difficult to identify the bulk composition of the planet from mass and radius alone (\citealt{Adams2008}, \citealt{Figueira2009}. Only spectroscopy of the exoplanet can hope to reveal its atmospheric composition unambiguously.

\subsubsection{TOI-2445}

TOI-2445 (J02531581+0003087) is an M dwarf situated at a distance of approximately 48.58 parsecs, with a mass of M=0.25$M_{\odot}$ and a radius of R=0.27$R_{\odot}$ (\citealt{Giacalone2022}). 
According to \cite{Giacalone2022}, TOI-2445 has $T_{\rm eff}$=3333$\pm157 K$, while our determination for the effective temperature is 3318$\pm 100K$, in very good agreement.  We obtain log $g$=4.94$\pm 0.13$ dex, while our results for the star's mass and radius are M=0.240$\pm 0.024 M_{\odot}$ and R=0.262$\pm 0.08 R_{\odot}$, respectively. 

This star hosts at least one known exoplanet, TOI-2445 b, that is categorized as a \textbf{s}uper-Earth, while JWST is observing its transit, eclipse, and phase curve in GO program 3784 (\citealt{Zhang2023}). Concerning its physical characteristics, TOI-2445 b has a mass of about 2.0$M_{\earth}$ and a radius of 1.25$\pm 0.08$ $R_{\earth}$ (\citealt{Giacalone2022}). Orbiting at only 0.006 AU, TOI-2445 b has an orbital period of $0.371$ days. 
Due to its small orbital radius, TOI-2445 b is subjected to intense irradiation, creating an extreme environment. 
TOI-2445 b's orbital eccentricity is assumed to be 0.0 (\citealt{Morello2023}) and using the semi-major axis of the orbit provided by \cite{Giacalone2022}, along with our results for the star's effective temperature and radius, we find a planetary equilibrium temperature of 936$\pm 28K$, while \cite{Giacalone2022} calculated a slightly higher equilibrium temperature of 1060$\pm 54 K$. 
In addition, we determined a planetary radius of 1.375$\pm 0.005 R_{\earth}$, slightly higher than that presented in the literature.

TOI-2445 has very few metallicity studies in the literature, with APOGEE and LAMOST reporting values of -0.44 and -0.54 dex, respectively. Our analysis yields [Fe/H]=-0.16$\pm 0.094$ dex, indicating a slightly sub-solar metallicity. This value is larger by $\sim$+0.3 to +0.4 dex from those obtained by APOGEE and LAMOST. 
This metallicity difference for such a cool M dwarf is not surprising. As discussed earlier in Section 4.1.1, the APOGEE DR 17 pipeline abundances are systematically lower (by about this same amount) than those derived from the spectroscopic analyses of \cite{Souto2021} and \cite{Wanderley2023}.  The similar discrepancy with the LAMOST metallicity likely results from the difficulties in measuring chemical abundances in cool M dwarfs from low-resolution optical spectra.
Using IRD spectra, \citealt{Morello2023} derived consistent metallicity values from spectral synthesis ([M/H]=-0.32$\pm$0.07), and SED fitting ([M/H] = -0.34$\pm$0.07) for TOI-2445. These values are located between our results and those from APOGEE and LAMOST.

TOI-2445 b's mass has not been observationally derived because it has yet to be confirmed via radial velocity (or by any other exoplanet discovery method). However, by assuming the exoplanet shares the same abundance ratios as its host star, we can determine the mass of TOI-2445 b using the ExoPlex code, an open-source mass-radius-composition solver.
ExoPlex calculates a planetary mass or radius based on its bulk composition of Mg, Si, Al, Ca, and Fe (\citealt{Unterborn2017}). The input parameters are our derived exoplanetary radius (1.375$\pm$0.005 $R_{\earth}$), along with our abundance ratios of Ca/Mg = 0.0638, Al/Mg = 0.0756, and Fe/Mg = 0.8958.  We were not able to determine a Si abundance for TOI-2445. Therefore, we adopt an alpha-elemental abundance as a proxy for Si, where [Si/Fe] $\sim$ [(O+Mg+Ca)/3Fe], in which we obtain Si/Mg = 0.9016. From these parameters, ExoPlex reports TOI-2445 b has a mass of 3.276$^{+0.448}_{-0.419}$ Earth masses. The errors are estimated by propagating the uncertainties in our derived abundances and those from the exoplanetary radius.  We find our mass determination of TOI-2445 b to be similar, within the uncertainties, to the one derived from \cite{Giacalone2022} using an estimate of the semi-amplitude of the radial velocity (RV) signal for this planet of $K_{\rm RV} = 4.5^{+2.8}_{-1.7}$ m.s$^{-1}$, which corresponds to a \textbf{$M_{\rm p} = 2.0^{+1.2}_{-0.7}$ $M_{\earth}$.}
The density of TOI-2445 b is 6.793$^{+0.005}_{-0.099}$ g.cm$^{-3}$, indicating the planet is likely to be composed of a rocky (Fe+Ni) core. 
Assuming \cite{Zeng2016} planetary density curves, we obtain TOI-2445 b to be at a very similar density distribution as the Earth, with a rock-dominated internal structure.

The proportions of a planet's core and mantle depend on complex factors, such as the amount of iron compared to the other constituent elements and the oxidation state of the core and the mantle. In ExoPlex, the ratios that most significantly contribute to a planet's core mass fraction are Fe/Mg and Si/Mg. Consequently, variations in these abundances result in a range of values for the core mass fraction, directly affecting the planetary mass and surface gravity. 
In the case of TOI-2445 b, we obtain a core mass fraction of 32.85$_{-0.049}^{+0.028}$, which is very similar to the Earth's (0.33) within the uncertainties. This modeling is consistent with what is expected, given the variations in Si/Mg and Fe/Mg abundances between the Earth and TOI-2445 b.

\subsubsection{GJ 3470}

Gliese 3470 is an M dwarf located at a distance of approximately 29.42 pc %from Earth 
(\citealt{Bonfils2012}). Our results for the stellar parameters and metallicity of GJ 3470 are $T_{\rm eff}$=3640$\pm$100K, log $g$=4.79$\pm$0.13dex, and [Fe/H]= +0.25$\pm$0.15dex. The results in the literature for this star agree very well with our determination. According to \cite{Awiphan2016}, GJ 3470 has $T_{\rm eff}$=3600$\pm$100K, log $g$=4.695$\pm$0.046dex, and [Fe/H]=+0.20$\pm$0.10dex, and \cite{Kosiarek2019} reports $T_{\rm eff}$=3652$\pm$50K, log $g$=4.658$\pm$0.035dex, and the same value for metallicity. 
The stellar mass obtained in this study for GJ 3470 was 0.455$\pm 0.046 M_{\odot}$, whereas for the stellar radius, our result is 0.492$\pm0.015 R_{\odot}$. 
As a comparison, \cite{Awiphan2016} determine $M_{\star}$=0.539$\pm$0.045$M_{\odot}$ and $R_{\star}$=0.547$\pm$0.018$R_{\odot}$.

GJ 3470 hosts at least one exoplanet, GJ 3470 b (\citealt{Bonfils2012}). 
The Neptune-like planet is located at 0.036 AU from its star, with an orbital period of approximately 3.337 days (\citealt{Awiphan2016}). The analysis of GJ 3470 b from \cite{Kosiarek2019}, presents an equilibrium temperature of 615$\pm$16K, with values for mass and radius of 12.58$_{-1.28}^{+1.31}$ $M_{\earth}$ and 3.88$\pm$0.32 $R_{\earth}$, respectively.

Using the semi-major axis of the orbit provided by \cite{Awiphan2016} and our determination for the effective temperature and stellar radius, the calculated equilibrium temperature for the exoplanet is 592$\pm$16$K$. Finally, the exoplanetary radius obtained is 4.352$\pm 0.028 R_{\earth}$, in addition to a planet mass of 11.276$\pm 0.497 M_{\earth}$.

The low density of GJ 3470 b would support the presence of an atmosphere dominated by H$_{2}$ (\citealt{Biddle2014}). \cite{Fukui2013} and \cite{Nascimbeni2013} observed wavelength-dependent variations in transit depths, indicating changes in atmospheric opacity. It is important to emphasize that \cite{Fukui2013} argued that GJ 3470 b would not have a thick layer of clouds. On the other hand, in the K-band, \cite{Crossfield2013} detected a flat transmission spectrum, implying a hazy, low-methane, and/or metal-rich atmosphere. \cite{Nascimbeni2013} found a difference in transit depth between ultraviolet and optical wavelengths, suggesting a Rayleigh-scattering slope consistent with a hazy atmosphere. Lastly, \cite{BourrierA2018} determined the presence of an extensive exosphere of neutral hydrogen around the planet. Using HST/WFC3 and {\em Spitzer} transits and eclipses, \cite{Benneke2019} reported an atmosphere with H$_2$O absorption, revealing low-to-moderate metallicity --- O/H = 0.2--18.0 relative to Solar --- and scattering clouds.

\subsection{The El/Fe abundance ratios}

The metallicities of the four planet-hosting stars in this study range from slightly metal-poor ([Fe/H]=-0.16) to metal-rich ([Fe/H]=+0.25). Their measured [el/Fe] ratios for the studied elements Na, Mg, Al, Si, K, Ca, Ti, V, and Mn (Table \ref{tab:abundances}) generally overlap with previous [El/Fe] vs [Fe/H] results for benchmark M dwarf stars from the studies of \cite{Souto2022} and the chemical evolution of these elements in the restricted range in metallicity of the sample stars (see figure 6 and 7 from \citealt{Souto2022}). The abundance ratio of C/O will be discussed next.

\subsection{The C/O abundance ratio}

%--------------------------------------------
\begin{figure*}
  \includegraphics[width=.95\textwidth]{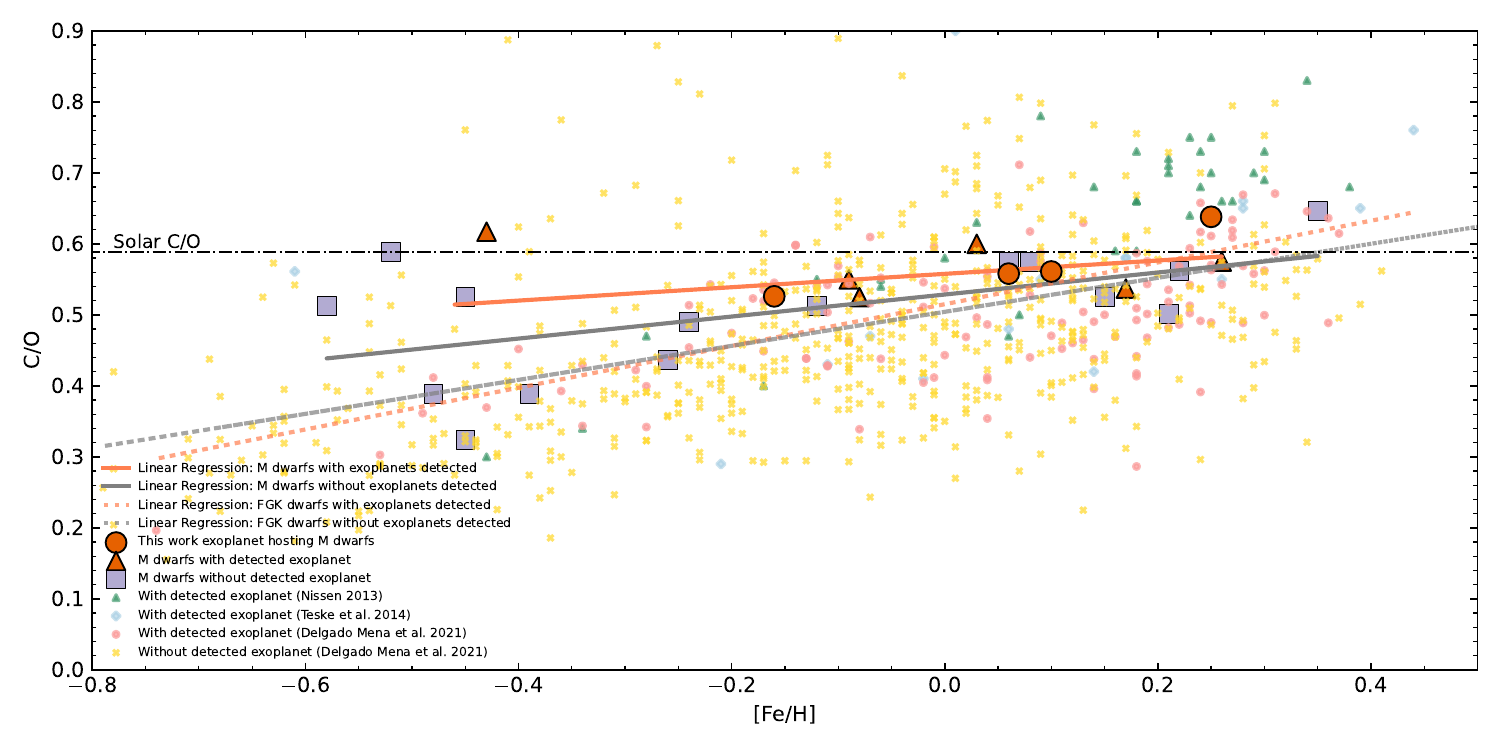}
  \caption{The [Fe/H] \textit{versus} C/O distribution for this work sample, as well as other M dwarfs studied by \cite{Souto2017}, \cite{Souto2018Ross}, \cite{Souto2020}. The orange circles represent the results for the four JWST targets studied here; Results for M dwarfs from our previous studies (all based on the same methodology as here) are also shown: orange triangles are M dwarfs with exoplanets detected, while the gray squares are M dwarfs without exoplanets detected so far. 
  We also present results for FGK dwarfs from \cite{Nissen2013} (green triangles; all with detected exoplanets), \cite{Teske2014} with (light blue diamonds) detected exoplanets, and \cite{DelgadoMena2021} with (pink circles), and without detected exoplanets (yellow cross). 
  Additionally, four curves represent the linear regressions of the [Fe/H] \textit{vs.} C/O distribution for M dwarfs with planet detected (solid orange line), M dwarfs without planet detected (solid gray line), FGK dwarfs with detected planets (dashed orange line), and FGK dwarfs without detected planets (dashed gray line).}
  \label{fig:C/O}
\end{figure*}
%--------------------------------------------

The C/O ratio can influence the availability of carbon and oxygen within exoplanet atmospheres, potentially causing them to become predominantly incorporated into molecules such as CH$_{4}$, H$_{2}$O, and CO. This phenomenon makes the carbon and oxygen content diagnostic of (and fundamentally shaping) the exoplanet temperature structure and composition (\citealt{Madhusudhan2012, Teske2014}). Similarly, the ratio of volatiles such as C and O to S has also been suggested as a complementary diagnostic of a planet's formation location and accretion history (\citealt{Pacetti2022,Schneider2021,Crossfield2023}).
Unfortunately, the S I lines are either blended or too weak to be properly studied for M dwarf stars in the APOGEE wavelength.

For the four studied stars, we obtain C/O=0.558$\pm$0.027 for TOI-1685, C/O=0.526$\pm$0.029 for TOI-2445, C/O=0.561$\pm$0.097 for GJ 436, and for GJ 3470 the C/O is 0.638$\pm$0.015 (Table 2).
We find that our C/O abundance ratio is smaller than solar for three stars and higher for the most metal-rich star, GJ 3470. The solar value reference adopted is from \cite{Asplund2021}, where C/O=0.589. 

In Figure \ref{fig:C/O}, we show the C/O ratio distribution as a function of metallicity for the four JWST targets studied here, along with other M dwarfs studied by \cite{Souto2017}, \cite{Souto2018Ross}, \cite{Souto2020}, which used the same analysis methodology as in this study. The orange circles represent the studied stars (all having detected exoplanets); orange triangles are M dwarfs studied by Souto et al. with exoplanets detected, while the gray squares are M dwarfs from Souto et al. without exoplanets detected so far.
We also present in this figure the C/O \textit{vs.} [Fe/H] of exoplanet hosting FGK dwarfs from \cite{Nissen2013} (green triangles), stars from \cite{Teske2014} (with detected exoplanets shown as a light blue diamond), and from \cite{DelgadoMena2021} (with detected exoplanets represented by a pink circle, and without detected exoplanets as a yellow cross). 

To explore C/O trends as a function of metallicity for hosting and non-hosting exoplanets, we display four curves representing the linear regressions of the [Fe/H] \textit{vs.} C/O distribution for M dwarfs with planet detected (solid orange line), M dwarfs without planet detected (solid gray line), FGK dwarfs with planet detected (dashed orange line), and FGK dwarfs without planet detected (dashed gray line).
We find no clear distinction or offset between the regressions for FGK dwarfs (with/without exoplanet detected to date) besides an intersection and in the metallicity about -0.20.
However, an offset is observed in the distribution of M dwarfs with and without exoplanets detected to date. The C/O ratio for M dwarfs hosting exoplanets is consistently higher by 0.01 to 0.05 dex compared to those without detected exoplanets. This difference is particularly pronounced in the metal-poor regime.
Given that the typical uncertainty in our C/O measurement is about 0.03 (see Table \ref{tab:abundances}), this offset resides at the limits of a statistically significant claim that exoplanets hosting M dwarfs display a systematic higher C/O abundance ratio.

% , and the host exoplanet stars displaying slightly higher C/O for the metal-rich sample ($\sim$0.02), while for the metal-poor the C/O is slightly lower ($\sim$-0.03).
% However, the M dwarf distribution displays a statistically significant offset where the C/O ratio for exoplanet hosting M dwarfs is roughly 0.05--0.08 higher than those without exoplanets detected to date. The typical uncertainty in our C/O measurement is about 0.03.

% In addition, we show four curves representing the linear regressions of the [Fe/H] \textit{vs.} C/O distribution for M dwarfs with planet detected (solid orange line), M dwarfs without planet detected (solid gray line), FGK dwarfs with planet detected (dashed orange line), and FGK dwarfs without planet detected (dashed gray line).
% There is no clear distinction or offset between the regressions for FGK dwarfs (with/without exoplanet detected to date) besides an intersection and in the metallicity about -0.20, and the host exoplanet stars displaying slightly higher C/O for the metal-rich sample ($\sim$0.02), while for the metal-poor the C/O is slightly lower ($\sim$-0.03). 
% However, the M dwarf distribution displays a statistically significant offset where the C/O ratio for exoplanet hosting M dwarfs is roughly 0.05--0.08 higher than those without exoplanets detected to date. The typical uncertainty in our C/O measurement is about 0.03.

We note that there is a distinct angular coefficient in the linear regression of the [Fe/H] \textit{vs.} C/O distribution for M dwarfs compared to FGK dwarfs. 
The slope for FGK dwarfs ($\sim$0.27) is roughly double that observed for M dwarfs ($\sim$0.14). Interestingly, there is no significant difference in slope when comparing the C/O distribution with and without detected exoplanets.
As discussed in \cite{Souto2020}, elemental abundance results for M dwarfs (obtained using the same methodology as here) and those for FGK stars from optical studies in the literature for wide-binary systems featuring a warmer primary and a secondary M dwarf showed no discernible systematic offset between the abundances. 
This substantiates the likelihood that the shallower slope observed in the M dwarf C/O \textit{vs.} [Fe/H] distribution is a genuine signature.

One possible physical explanation for the offset in the C/O distribution could be attributed to the atomic diffusion process, which depletes abundances in FG dwarfs, particularly intensifying as stellar metallicity decreases (Souto et al., 2018, 2019, 2021; Dotter, 2017). However, when studying abundance ratios like C/O, we anticipate that atomic diffusion signatures may be mitigated due to their similar intensity across different elements. Other contributing factors may include variations in age distribution or the transient nature of many M dwarfs in the solar neighborhood, possibly originating from different regions of our galaxy. This is supported by the observation that a significant portion of M dwarfs within 100 pc of the Sun exhibit higher proper motions compared to AFG dwarfs (RECONS catalog; www.recons.org). In future investigations, we aim to explore potential correlations between these variations and physical processes in Galactic chemical evolution.

\section{Summary}

We conducted a detailed spectroscopic analysis, using SDSS APOGEE $H$-band spectra, for four M-dwarf stars, TOI-1685, GJ 436, TOI-2445, and GJ 3470. These stars host Earth-like or Neptune-like exoplanets and are scheduled for observation by JWST. 
Accurate spectroscopic characterization of the stellar parameters and individual abundances is crucial for enhancing the precision of transmission spectroscopic observations of their associated exoplanets. Notably, our spectral analysis did not reveal any significant Zeeman split lines, which would suggest, in principle, that potential interference from starspots or stellar activity during exoplanet transits may be minimal.

We followed the same analysis methodology discussed in our previous works \cite{Souto2020} to derive stellar effective temperatures and surface gravities from a spectral synthesis of water and OH lines present in the APOGEE spectra. The derived sample has effective temperatures in the range between $\sim$3300 - 3600 K and log g between $\sim$ 4.6 - 4.9. 
We also derived stellar masses and radii for the sample using calibrations from our previous work \cite{Souto2020}, resulting in stellar radii ranging from 0.262 to 0.492 $R_{\odot}$, and stellar masses ranging from 0.240 to 0.455 $M_{\odot}$. 

Planetary masses and radii were also derived for the studied sample using the stellar parameters from this work. 
Using equation 3, we find that TOI-1685 b, GJ 436 b, TOI-2445 b, and GJ 3470 b have $R$/$R$$_{\earth}$ of 1.759, 3.659, 1.385, and 4.352. Using equation 4, we find that
and for TOI-1685 b, GJ 436 b, and GJ 3470 b have masses of $M$/$M$$_{\earth}$ of 3.142, 18.606, and 11.276, respectively.

The metallicities obtained from the four planet-hosting M dwarf stars range from slightly metal-poor to metal-rich. For TOI-1685, we obtained a near-solar metallicity, with [Fe/H]=0.06$\pm$0.18 dex. In the case of GJ 436, we measured a relatively higher iron abundance of [Fe/H]=0.10$\pm$0.20 dex, while for TOI-2445, we derived a sub-solar metallicity of [Fe/H]=-0.16$\pm$0.094 dex. Finally, our analysis indicated that GJ 3470 is the most metal-rich star, with [Fe/H]=0.25$\pm$0.15.
The measured abundance ratios of [Na/, Mg/, Al/, Si/, K/, Ca/, Ti/, V/, Mn/ Fe] in the four stars generally overlapped with the chemical evolution of these elements in the Galactic disk within the restricted range in the metallicity of the sample stars.

Based on our derived abundance ratios of Ca/Mg, Si/Mg, Al/Mg, and Fe/Mg, we determined a mass of 3.276$^{+0.448}_{-0.419}$ $M_{\earth}$ for TOI-2445 b, and a density of 6.793$^{+0.005}_{-0.099}$ g.cm$^{-3}$, having a rock-dominated internal structure likely similar to the Earth (core mass fraction of 0.329$_{-0.049}^{+0.028}$).

To constrain the chemical properties of the protoplanetary disks that gave rise to these exoplanets, we determined the fundamental carbon-to-oxygen (C/O) ratios for the exoplanet-hosting M dwarfs in this small sample, revealing C/O ratios of 0.558$\pm$0.027 for TOI-1685, 0.526$\pm$0.029 for TOI-2445, 0.561$\pm$0.097 for GJ 436, and 0.638$\pm$0.015 for GJ 3470.  These C/O ratios (for four M dwarfs), when combined with our previous abundance analyses of M dwarfs in \citet{Souto2017,Souto2018Ross,Souto2020}, yield a sample of 28 M dwarfs whose values for C/O were derived from homogeneous analyses of APOGEE spectra.
The results from this combined sample of M dwarfs, which consists of both exoplanet hosting and non-hosting stars, reveal a consistent trend among the M dwarfs, where the C/O ratios for exoplanet-hosting stars are consistently higher, by about 0.01 to 0.05, when compared to M dwarfs without detected exoplanets (Figure \ref{fig:C/O}). This offset is not observed for FGK dwarfs.
This result could suggests that among M dwarfs, a protoplanetary disk richer in carbon relative to oxygen is more likely to form super-Earth or mini-Neptune exoplanets.
Also, we find that the linear regression from the [Fe/H] \textit{vs.} C/O distribution reveals a notable difference in the angular coefficient for M dwarfs compared to FGK dwarfs. FGK dwarfs exhibit a slope of approximately 0.27, roughly twice that of M dwarfs, which show a slope of around 0.13. In addition, there is no significant difference in slope when comparing the C/O distribution with and without detected exoplanets for the FGKM dwarfs.

Finally, this paper presents (in an appendix) an uncertainty atlas for the abundances of M dwarfs with effective temperatures between 3200 - 4000 K, obtained from computed spectra between 1.5 - 1.7$\mu$m, compiling abundance sensitivities and uncertainties stemming from atmospheric parameter ($T_{\rm eff}$, log $g$, [Fe/H], $\xi$) variations, SNR, and pseudo-continuum placements. This atlas can enable a more realistic estimation of abundance uncertainties for M dwarf stars studied in the $H$-band.

\begin{acknowledgments}
We thank the referee for suggestions that helped improve the paper. D.S. thanks the National Council for Scientific and Technological Development – CNPq.
N.H. acknowledges support from NSF AAG grant No. 2108686, and from NASA ICAR grant No. NNH19ZDA001N.
R.L. acknowledges funding from University of La Laguna through the Margarita Salas Fellowship from the Spanish Ministry of Universities ref. UNI/551/2021-May 26, and under the EU Next Generation funds. This study was financed in part by the Coordenação de Aperfeiçoamento de Pessoal de Nível Superior – Brasil (CAPES) – Finance Code 001. 

Funding for the Sloan Digital Sky Survey IV has been provided by the Alfred P. Sloan Foundation, the U.S. Department of Energy Office of Science, and the Participating Institutions. SDSS-IV acknowledges support and resources from the Center for High-Performance Computing at the University of Utah. The SDSS website is www.sdss.org.
SDSS-IV is managed by the Astrophysical Research consortium for the Participating Institutions of the SDSS Collaboration including the Brazilian Participation Group, the Carnegie Institution for Science, Carnegie Mellon University, the Chilean Participation Group, the French Participation Group, Harvard-Smithsonian Center for Astrophysics, Instituto de Astrof\'isica de Canarias, The Johns Hopkins University, 
Kavli Institute for the Physics and Mathematics of the Universe (IPMU) /  University of Tokyo, Lawrence Berkeley National Laboratory, Leibniz Institut f\"ur Astrophysik Potsdam (AIP),  Max-Planck-Institut f\"ur Astronomie (MPIA Heidelberg), Max-Planck-Institut f\"ur Astrophysik (MPA Garching), Max-Planck-Institut f\"ur Extraterrestrische Physik (MPE), National Astronomical Observatory of China, New Mexico State University, New York University, University of Notre Dame, Observat\'orio Nacional / MCTI, The Ohio State University, Pennsylvania State University, Shanghai Astronomical Observatory, United Kingdom Participation Group,
Universidad Nacional Aut\'onoma de M\'exico, University of Arizona, University of Colorado Boulder, University of Oxford, University of Portsmouth, University of Utah, University of Virginia, University of Washington, University of Wisconsin, Vanderbilt University, and Yale University.

\textit{\facility {Sloan Digital Sky Survey (SDSS-IV)}}

\software{Turbospectrum (\citealt{AlvarezPlez1998}, \citealt{Plez2012}), MARCS (\citealt{Gustafsson2008}), Bacchus (\citealt{Masseron2016}), ExoPlex (\citealt{Unterborn2017}), Matplotlib (\citealt{Hunter2007}), Numpy (\citealt{Vanderwalt2011}), RECONS catalog; www.recons.org.}
\end{acknowledgments}

\vspace{5mm}

\appendix
\section{An uncertainty Atlas in M dwarf $H$-band spectra}

Abundance determinations are based on modeling the stellar atmospheres, which involves understanding how radiation interacts with a star's outer layers, and the precision of these models plays a vital role in the accuracy of abundance calculations. Any uncertainties in the stellar parameters, such as effective temperature, surface gravity (log $g$), and microturbulence ($\xi$), directly affect abundance estimates. Departures from local thermodynamic equilibrium (non-LTE effects) can also affect the line strengths and shapes and introduce additional complexities into the analyses. In addition, spectroscopic observations are not immune to instrumental and observational errors. Variability in telescope performance, detector sensitivity, and data reduction techniques can introduce systematic errors. These errors can be particularly challenging to quantify accurately. 
In addition to systematic uncertainties, abundance determinations also involve random errors. These errors are associated with the inherent randomness in observations and measurements. They can include statistical fluctuations and variations due to atmospheric, instrumental, and/or line selection conditions. Consequently, noise and measurement errors may affect abundance determinations. In particular, weak lines are more susceptible to these sources of uncertainty than stronger lines. 

This appendix presents the uncertainties in the derived abundances but also provides an uncertainty atlas based on the sensitivities of the abundances for baseline models with effective temperatures varying from 3200 - 3900 K in steps of 100 K. Our analysis of the uncertainties encompasses changes in atmospheric parameters ($T_{\rm eff}$, log $g$, [Fe/H], and microturbulence), pseudo-continuum location, and signal-to-noise ratios. The selected wavelength and resolution range are based on the capability of the APOGEE spectrograph (but can be extended for higher-resolution studies), which is widely used for M dwarf characterization and is integral to the SDSS-V MWM survey. 
Additional sources of uncertainty in our abundance analysis were not extensively addressed in this study, as they are generally expected to have a smaller impact than atmospheric parameter variations, SNR, and pseudo-continuum displacements. For instance, blends resulting from H$_{2}$O and FeH opacities, as observed in Figure \ref{fig:geral}, can contribute to pseudo-continuum depletion and potentially affect other elemental abundances. However, previous work by \cite{Souto2017} suggests that their influence is minor.

\subsection{Uncertainties due to changes in the Atmospheric parameters} \label{parameters}

Variations in $T_{\rm eff}$ (Figure \ref{fig:geral} top panel) span from $4000$ to $3200K$ in regular increments of $100K$ (color-coded from blue to red in the Figure). As $T_{\rm eff}$ decreases, the spectral lines become less prominent, presenting challenges for analysis, a trend we observe in other wavelengths in the $H$-band as well.
Among the displayed lines, the Mg I lines exhibit the highest sensitivity to changes in $T_{\rm eff}$, while the depth of OH lines remains relatively constant at approximately 0.75 in normalized synthetic spectra. 
Notably, this panel highlights the presence of a pseudo-continuum depletion. 
As $T_{\rm eff}$ decreases, the opacity of H$_{2}$O transitions intensifies, leading to continuum suppression. This effect becomes more pronounced below 3500K, resulting in a typical depletion of approximately 15 to 20\% from the continuum. It's worth noting that this depletion introduces significant uncertainties related to the pseudo-continuum definition (refer to Section \ref{continumm}).

For Figure \ref{fig:geral} middle panel, we vary model metallicity from $-1.0$ to $0.5$ at regular intervals of $0.25$ dex, keeping all abundances constant at the solar scale. 
We assume the $T_{\rm eff}$ fixed at $3500K$, while surface gravity remains at $4.80$ dex. We observe a difference in the line profile much smaller than changing $T_{\rm eff}$. The Mg I lines are more sensitive to model metallicity variations, whereas the OH lines' depth remains relatively unchanged. Additionally, the continuum shows no significant variation across the different ranges of model metallicity studied. 
% We note that the but there is a stronger pseudo-continuum displacement for those metal-rich models due to the presence of more O to be bound in H$_{2}$O molecule and therefore suppressing the continuum. 
In Figure \ref{fig:geral} bottom panel, we present synthesis with changes in log $g$ ranging from $4.0$ to $5.4$ dex at regular intervals of $0.20$ dex. The effective temperature and metallicity are fixed at $3500K$ and solar value, respectively. 
We note that lower values for surface gravity are linked to deeper spectral lines, which gradually smooth out as this atmospheric parameter increases with higher log $g$ values. It is related to increasing internal pressure and higher rates of collisional broadening in the line formation.
The behavior is similar for all species in the window and is reminiscent of the entire $H$-band spectra.

We initially constructed a spectral synthesis grid (assumed to match APOGEE resolution) with fixed atmospheric parameter values to assess the impact of errors in abundance measurements due to atmospheric parameter uncertainties. This grid is, therefore, interpreted as our simulated observed spectra. 
Subsequently, we calculate individual abundances based on these simulated spectra. Next, we systematically vary the model atmosphere parameters: effective temperature by $\pm$100 K and $\pm$50 K, log $g$ and [Fe/H] by $\pm$0.20 and $\pm$0.10 dex, and microturbulence velocity by $\pm$0.50 and $\pm$0.25 km.s$^{-1}$. We then recompute the corresponding abundances, assuming these changes in the model atmosphere.
The discrepancy between the abundances obtained from the synthesis grid and those resulting from the variation in atmospheric parameters defines our typical uncertainty attributable to changes in atmospheric parameters ($\sigma_{\rm AP}$).

In Table \ref{tab:sensitivity}, we display the abundance sensitivities as a function of changes in stellar atmospheric parameters. The elements we study are Fe (from 
Fe I and FeH lines), C (from CO lines), O (from OH and H$_{2}$O lines), and the elements Na, Mg, Al, Si, K, Ca, Ti, V, Cr, Mn, and Ni, all from neutral lines. The sensitivity analysis adopts baseline models with effective temperature values ranging from $3900K$ to $3200K$ at regular intervals of $100 K$. We expect these respective abundance sensitivity shifts to be linear (presented in Table \ref{tab:sensitivity}), and we can interpolate them whether necessary to obtain an abundance sensitivity for a different set of uncertainties in the atmospheric parameters. 
It's worth noting that as the reference effective temperatures decrease, some spectral lines become impractical to investigate properly due to molecular blends. This is why there are gaps in Table \ref{tab:sensitivity}. Chromium and Nickel already become impractical for analysis at effective temperatures of about $3600$ K. Vanadium and Manganese, on the other hand, become unfeasible for investigation only from effective temperatures of around $3400 K$. 

The elements Mg and Si are the ones displaying the highest abundance sensitive to changes in $T_{\rm eff}$ ($\ga$ 0.20 dex), followed by Fe (from FeI lines), O (from H$_{2}$O lines), and Al ($\sim$ 0.10 dex). The ones showing less abundance sensitivity to $T_{\rm eff}$ are C and O (from OH lines) ($\la$ 0.03 dex). 
Regarding changes in log $g$, Mg, Si, and Fe (Fe I) are the most responsive, with variations of approximately 0.15 dex. On the other hand, C and O (H$_{2}$O) exhibit lower sensitivity to log $g$.
The abundance sensitivity to changes in metallicity is relatively consistent across all elements. O (from H$_{2}$O lines), Na and Fe (from FeH lines) are typically the most sensitive, with changes exceeding 0.10 dex.
Abundance deviations from changes in microturbulence velocity are minimal for all elements, typically staying below 0.03 dex.

%--------------------------------------------
\begin{figure*}
  \centering
  \includegraphics[width=1.0\textwidth]{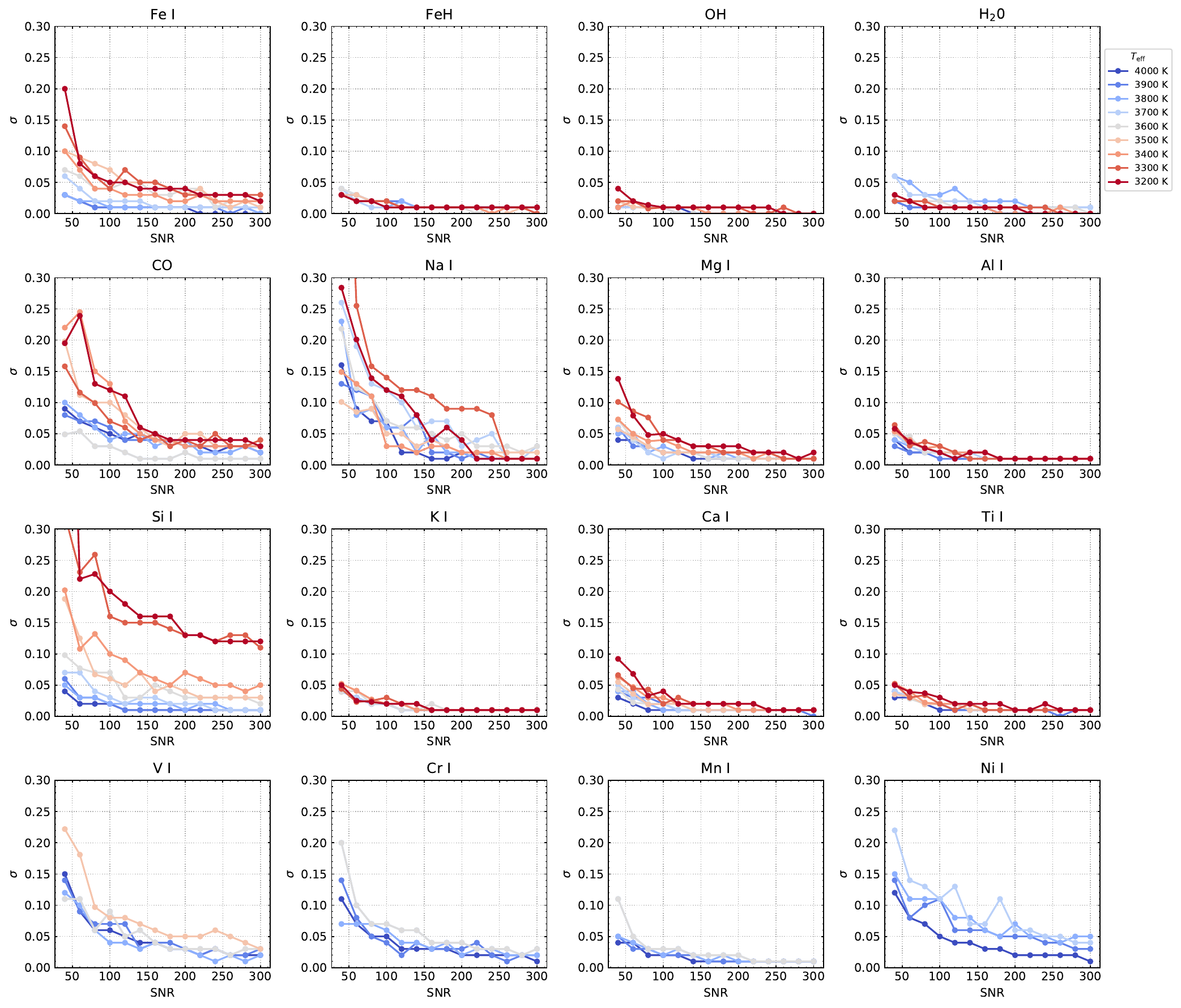}\label{fig:a} \\
  \caption{Signal-to-noise (SNR) versus uncertainties due to changes in the signal-to-noise ratio ($\sigma_{\rm SNR}$). }
  \label{fig:sigma}
\end{figure*}
%--------------------------------------------

\subsection{Uncertainties due to changes in the signal-to-noise ratio} \label{snr}

The signal-to-noise per pixel ratio (SNR) is a measure used to assess the quality or strength of a signal in relation to the background noise level. A higher SNR indicates a stronger, more reliable signal than the noise, making it easier to detect and analyze the data. In general, high-resolution stellar spectroscopy assumes SNR $>$ 60 -- 100 \citep{Jofre2019} to be enough to determine individual abundances precisely. In this paper case, we will validate these for M dwarfs in the $H$-band.

We apply a normal random distribution (using the Python library Numpy) to introduce noise into our spectral synthesis. Therefore, we obtain a synthesis with a controlled degree of noise where we assume them as our ``observed spectra''. Subsequently, we add noise levels ranging from SNR of 40 to 300 in steps of 20, and we apply this noise for all ranges of $T_{\rm eff}$ adopted in this work ($T_{\rm eff}$ of 3200 to 4000K, in 100K steps). We assume fix values of log $g$ =  4.80 dex, [Fe/H] = 0.00 dex, and $\xi$ = 1.00 km.s$^{-1}$ in this case.

The noise added accounts for a random variation of each pixel in our spectral synthesis. To obtain the abundance uncertainty relative to changes in SNR, we then create a total of 20 syntheses for each set of atmospheric parameters ($T_{\rm eff}$, log $g$, $\xi$, and [Fe/H]) and SNR analyzed, and compute their respective elemental abundances. We assume the SNR uncertainty ($\sigma_{\rm SNR}$) to be the standard deviation of the mean of the abundances derived from these 20 random ``observed spectra''. 
We adopt a number of 20 interactions because both mean and standard deviation reach a stable plateau, obviating the necessity for further iterations.

In Table \ref{tab:noise}, we provide an overview of uncertainties related to changes in SNR. As our highest SNR value studied is 300, we assume higher SNR stars to have the same abundance uncertainties as those with SNR = 300.
Figure \ref{fig:sigma} illustrates the relationship between SNR and uncertainty ($\sigma$), with each element along with the set of the studied $T_{\rm eff's}$ represented in separate panels. As expected, the figure demonstrates a general trend of decreasing standard deviation of the mean ($\sigma_{\rm SNR}$) as SNR increases.
Regarding sensitivity to $T_{\rm eff}$, we typically observe higher uncertainties at fixed SNR values for lower $T_{\rm eff}$. 
Notably, FeH and OH lines consistently exhibit uncertainties below 0.05 dex, owing to their numerous well-defined lines (refer to Table \ref{tab:lines}). The same is true for K I lines, but only two lines are available in the APOGEE wavelength coverage.
Similar trends, with slightly higher uncertainties for SNR $<$ 100, are observed for H$_{2}$O, Al, K, Ca, Ti, and Mn. 
For Fe (Fe I lines), CO, Mg, V, and Ni, the uncertainty remains around 0.05 dex when SNR exceeds 200. However, lower SNR values, such as SNR = 40 for cool stars, can lead to uncertainties exceeding 0.15 dex.
Notably, Na and Si exhibit higher uncertainties due to changes in SNR. Even at SNR = 300, the uncertainty remains close to 0.05 dex. For Si, when $T_{\rm eff}$ is 3300K, $\sigma_{\rm SNR}$ consistently exceeds 0.10 dex regardless of the adopted SNR.

\subsection{Uncertainties due to changes in the pseudo-continuum definition} \label{continumm}

The complexity of M dwarf spectra poses challenges in defining the pseudo-continuum, which is expected to be transition-free. Consequently, uncertainties in pseudo-continuum definition can lead to shifts in the normalized spectral synthesis flux, influencing the measurement of abundances and introducing associated uncertainties. This section lists uncertainties due to pseudo-continuum shifts ($\sigma_{\rm PC}$). To evaluate uncertainties associated with variations in the pseudo-continuum, we utilize the spectral synthesis grid (discussed in Section \ref{parameters}) as our ``observed spectra'' and calculate individual line abundances for all studied elements. We introduce shifts of -2\%, -1\%, +1\%, and +2\% in the normalized flux of the ``observed spectra'' and then compute their abundances.
We consider a $\pm$2\% deviation in the normalized flux spectra as lower and upper limits of shifts. Changes beyond this range make determining abundances challenging (or impossible) for most species. We define our $\sigma_{\rm PC}$ as the difference between the individual abundance derived with and without the shift in the normalized flux spectra.

Table \ref{tab:displacement} summarizes uncertainties in pseudo-continuum displacement at temperatures ranging from $4000$ to $3200K$ for all studied species. Each row in the table corresponds to a specific element, while the columns represent percentage variations in pseudo-continuum displacement relative to the reference condition, including $\pm2$\% and $\pm1$\%. The values in each table cell indicate the abundance change for the respective element under these conditions. 

The uncertainties here are usually more significant for weaker lines. For Na, V, Cr, and Ni, for example, the uncertainty can be higher than 0.20 dex, assuming $\pm2$\% shifts. For most species, the $\sigma_{\rm PC}$ is smaller than 0.10 for shifts of $\pm1$\% in the pseudo-continuum definition.
These values are valuable for interpreting spectroscopic analyses and quantifying uncertainties, facilitating a more precise understanding of how pseudo-continuum displacement variations impact the abundance of different elements.

\clearpage
\startlongtable
\begin{longrotatetable}
\begin{deluxetable*}{lcccc|cccc|cccc|cccc}
\rotate
\tabletypesize{\scriptsize}
\tablecaption{Abundance sensitive to changes in stars' atmospheric parameters.}
\tablewidth{0pt}
\label{tab:sensitivity}
\tablehead{
\colhead{Element} &
\multicolumn{4}{c|}{$T_{\rm eff}$} & \multicolumn{4}{c|}{Log $g$} & \multicolumn{4}{c|}{Fe/H} & \multicolumn{4}{c}{$\xi$} \\
% \hline
\colhead{} &
\colhead{+100} & 
\colhead{+50} & 
\colhead{-50} & 
\colhead{-100} &
\colhead{+0.20} &
\colhead{+0.10} &
\colhead{-0.10} &
\colhead{-0.20} &
\colhead{+0.20} &
\colhead{+0.10} &
\colhead{-0.10} &
\colhead{-0.20} &
\colhead{+0.50} & 
\colhead{+0.25} & 
\colhead{-0.25} &
\colhead{-0.50} \\
\colhead{} &
\colhead{(K)} & 
\colhead{(K)} & 
\colhead{(K)} & 
\colhead{(K)} &
\colhead{(dex)} &
\colhead{(dex)} &
\colhead{(dex)} &
\colhead{(dex)} &
\colhead{(dex)} &
\colhead{(dex)} &
\colhead{(dex)} &
\colhead{(dex)} &
\colhead{(km.s$^{-1}$)} & 
\colhead{(km.s$^{-1}$)} & 
\colhead{(km.s$^{-1}$)} &
\colhead{(km.s$^{-1}$)} \\
}
\startdata
Assuming $T_{\rm eff}$ = 3900 K \\
Fe (FeI) & -0.17 & -0.09 & 0.09 & 0.18  & 0.18 & 0.09 & -0.09 & -0.17  & -0.01 & -0.01 & 0.03 & 0.05  & 0.00 & 0.00 & 0.00 & 0.00 \\
Fe (FeH) & 0.03 & 0.01 & -0.01 & -0.03  & 0.05 & 0.03 & -0.03 & -0.05  & 0.09 & 0.04 & -0.05 & -0.10  & -0.03 & -0.01 & 0.01 & 0.02 \\
C & 0.06 & 0.00 & 0.00 & 0.00  & 0.03 & 0.02 & -0.01 & -0.01  & 0.18 & 0.09 & -0.07 & 0.02  & 0.01 & 0.01 & 0.00 & 0.00 \\
O (H$_{2}$O) & 0.1 & 0.07 & -0.02 & -0.04  & 0.00 & 0.00 & 0.00 & 0.00  & 0.20 & 0.14 & -0.10 & -0.21  & 0.00 & 0.00 & 0.00 & 0.00 \\
O (OH) & 0.00 & 0.00 & -0.01 & -0.01  & 0.02 & 0.01 & -0.02 & -0.06  & 0.05 & 0.00 & -0.09 & -0.17  & -0.02 & -0.01 & 0.01 & 0.01 \\
Na & -0.20 & -0.02 & 0.00 & -0.01  & 0.15 & 0.06 & -0.04 & -0.27  & -0.13 & -0.18 & 0.07 & 0.11  & 0.00 & 0.00 & 0.01 & 0.00 \\
Mg & -0.30 & -0.14 & 0.12 & 0.22  & 0.17 & 0.09 & -0.11 & -0.24  & -0.11 & -0.05 & 0.03 & 0.03  & 0.00 & 0.00 & 0.00 & 0.00 \\
Al & -0.08 & -0.03 & 0.04 & 0.06  & 0.06 & 0.04 & -0.04 & -0.09  & -0.03 & -0.01 & 0.00 & -0.01  & 0.00 & 0.00 & 0.01 & 0.01 \\
Si & -0.24 & -0.13 & 0.12 & 0.22  & 0.21 & 0.11 & -0.12 & -0.23  & -0.05 & -0.03 & 0.03 & 0.04  & -0.01 & -0.01 & 0.00 & 0.00 \\
K & 0.00 & -0.01 & 0.00 & 0.01  & 0.05 & 0.02 & -0.03 & -0.06  & -0.02 & -0.02 & 0.02 & 0.04  & -0.01 & 0.00 & 0.00 & 0.00 \\
Ca & -0.03 & -0.02 & 0.01 & 0.02  & 0.09 & 0.05 & -0.06 & -0.13  & -0.08 & -0.04 & 0.03 & 0.07  & -0.01 & -0.01 & 0.00 & 0.00 \\
Ti & -0.01 & 0.00 & 0.02 & 0.02  & 0.09 & 0.05 & -0.04 & -0.08  & 0.02 & 0.01 & 0.01 & 0.00  & -0.01 & 0.00 & 0.01 & 0.02 \\
V & 0.04 & 0.02 & -0.01 & -0.03  & 0.13 & 0.08 & -0.04 & -0.08  & 0.09 & 0.03 & 0.02 & 0.01  & 0.00 & 0.00 & 0.00 & 0.00 \\
Cr & -0.11 & -0.06 & 0.08 & 0.15  & 0.17 & 0.09 & -0.08 & -0.13  & 0.01 & -0.01 & 0.06 & 0.08  & 0.00 & 0.00 & 0.00 & 0.00 \\
Mn & -0.14 & -0.07 & 0.08 & 0.16  & 0.17 & 0.09 & -0.08 & -0.15  & 0.01 & -0.01 & 0.04 & 0.07  & 0.00 & 0.00 & 0.00 & 0.00 \\
Ni & -0.06 & -0.01 & 0.21 & 0.26  & 0.43 & 0.29 & -0.14 & -0.27  & -0.15 & -0.08 & 0.27 & 0.29  & -0.01 & 0.00 & 0.01 & 0.01 \\
\hline
Assuming $T_{\rm eff}$ = 3800 K \\
Fe (FeI) & -0.16 & -0.09 & 0.11 & 0.19  & 0.19 & 0.10 & -0.09 & -0.17  & 0.00 & -0.01 & 0.04 & 0.07  & 0.00 & 0.00 & 0.00 & 0.00 \\
Fe (FeH) & 0.04 & 0.02 & -0.02 & -0.03  & 0.04 & 0.02 & -0.03 & -0.05  & 0.10 & 0.05 & -0.06 & -0.11  & -0.04 & -0.02 & 0.01 & 0.03 \\
C & 0.01 & 0.01 & -0.07 & -0.07  & 0.06 & 0.03 & -0.02 & -0.05  & 0.12 & 0.03 & -0.05 & -0.13  & -0.01 & 0.00 & 0.01 & 0.00 \\
O (H$_{2}$O) & 0.06 & 0.03 & -0.02 & -0.04  & 0.00 & 0.01 & 0.01 & 0.01  & 0.30 & 0.25 & -0.10 & -0.20  & 0.01 & 0.01 & 0.01 & 0.01 \\
O (OH) & 0.00 & 0.00 & 0.00 & 0.00  & 0.02 & 0.01 & -0.02 & -0.06  & 0.06 & 0.00 & -0.09 & -0.17  & -0.02 & -0.01 & 0.01 & 0.01 \\
Na & -0.15 & -0.13 & 0.06 & 0.07  & 0.18 & 0.10 & -0.17 & -0.20  & -0.04 & -0.10 & 0.13 & 0.14  & -0.13 & -0.13 & 0.00 & 0.00 \\
Mg & -0.26 & -0.13 & 0.13 & 0.23  & 0.17 & 0.09 & -0.10 & -0.20  & -0.09 & -0.04 & 0.04 & 0.05  & 0.00 & 0.00 & 0.00 & 0.00 \\
Al & -0.10 & -0.05 & 0.05 & 0.09  & 0.07 & 0.04 & -0.04 & -0.10  & -0.03 & -0.01 & 0.01 & 0.00  & 0.00 & 0.00 & 0.01 & 0.01 \\
Si & -0.24 & -0.12 & 0.15 & 0.27  & 0.23 & 0.12 & -0.11 & -0.22  & -0.05 & -0.02 & 0.06 & 0.09  & 0.00 & 0.00 & 0.01 & 0.01 \\
K & 0.02 & 0.01 & 0.00 & 0.00  & 0.04 & 0.02 & -0.03 & -0.05  & -0.02 & -0.01 & 0.02 & 0.04  & 0.00 & 0.00 & 0.00 & 0.00 \\
Ca & -0.03 & -0.02 & 0.02 & 0.03  & 0.10 & 0.05 & -0.06 & -0.12  & -0.08 & -0.03 & 0.04 & 0.08  & 0.00 & 0.00 & 0.00 & 0.00 \\
Ti & -0.02 & -0.02 & 0.01 & 0.02  & 0.08 & 0.04 & -0.04 & -0.08  & 0.02 & 0.01 & 0.00 & -0.01  & -0.02 & -0.01 & 0.01 & 0.02 \\
V & 0.05 & 0.03 & -0.02 & -0.03  & 0.13 & 0.07 & -0.04 & -0.08  & 0.09 & 0.04 & 0.02 & 0.02  & 0.00 & 0.00 & 0.00 & 0.00 \\
Cr & -0.10 & -0.05 & 0.08 & 0.19  & 0.16 & 0.09 & -0.06 & -0.11  & 0.03 & 0.01 & 0.05 & 0.08  & 0.00 & 0.00 & 0.00 & 0.00 \\
Mn & -0.12 & -0.07 & 0.09 & 0.16  & 0.17 & 0.09 & -0.08 & -0.15  & 0.02 & 0.00 & 0.05 & 0.08  & 0.00 & 0.00 & 0.00 & 0.00 \\
Ni & 0.08 & -0.04 & 0.14 & 0.15  & 0.35 & 0.24 & -0.19 & -0.34  & -0.13 & -0.16 & 0.25 & 0.22  & -0.11 & -0.07 & 0.02 & 0.05 \\
\hline
Assuming $T_{\rm eff}$ = 3700 K \\
Fe (FeI) & -0.11 & -0.06 & 0.09 & 0.20  & 0.19 & 0.10 & -0.08 & -0.15  & 0.03 & 0.01 & 0.06 & 0.10  & 0.00 & 0.00 & 0.00 & 0.00 \\
Fe (FeH) & 0.05 & 0.02 & -0.02 & -0.04  & 0.03 & 0.01 & -0.02 & -0.04  & 0.11 & 0.05 & -0.06 & -0.12  & -0.05 & -0.03 & 0.02 & 0.04 \\
C & 0.02 & 0.01 & -0.01 & 0.00  & 0.04 & 0.02 & -0.03 & -0.05  & 0.11 & 0.01 & -0.07 & -0.15  & 0.00 & 0.00 & -0.01 & -0.01 \\
O (H$_{2}$O) & 0.09 & 0.04 & -0.03 & -0.06  & -0.01 & -0.01 & 0.00 & 0.00  & 0.30 & 0.24 & -0.11 & -0.21  & 0.11 & -0.01 & 0.00 & 0.00 \\
O (OH) & 0.00 & 0.00 & 0.00 & 0.00  & 0.02 & 0.01 & -0.01 & -0.04  & 0.06 & 0.01 & -0.08 & -0.17  & -0.02 & -0.01 & 0.01 & 0.02 \\
Na & 0.01 & 0.00 & -0.01 & -0.02  & 0.12 & 0.03 & -0.02 & -0.04  & 0.24 & 0.08 & 0.07 & 0.16  & 0.00 & 0.00 & 0.00 & 0.00 \\
Mg & -0.24 & -0.13 & 0.12 & 0.22  & 0.15 & 0.08 & -0.08 & -0.17  & -0.06 & -0.03 & 0.03 & 0.04  & 0.00 & 0.00 & 0.00 & 0.00 \\
Al & -0.11 & -0.05 & 0.04 & 0.09  & 0.06 & 0.03 & -0.04 & -0.09  & -0.03 & -0.01 & 0.00 & -0.01  & -0.01 & -0.01 & 0.00 & 0.00 \\
Si & -0.27 & -0.14 & 0.13 & 0.27  & 0.23 & 0.13 & -0.12 & -0.23  & -0.08 & -0.04 & 0.07 & 0.09  & 0.00 & 0.00 & 0.01 & 0.01 \\
K & 0.05 & 0.03 & -0.01 & -0.01  & 0.03 & 0.02 & -0.01 & -0.02  & 0.00 & 0.00 & 0.02 & 0.03  & 0.00 & 0.00 & 0.01 & 0.01 \\
Ca & -0.04 & -0.01 & 0.02 & 0.04  & 0.10 & 0.05 & -0.06 & -0.12  & -0.07 & -0.03 & 0.04 & 0.08  & 0.00 & 0.00 & 0.00 & 0.00 \\
Ti & -0.02 & -0.01 & 0.01 & 0.02  & 0.07 & 0.04 & -0.04 & -0.08  & 0.04 & 0.01 & -0.01 & -0.03  & -0.02 & -0.01 & 0.01 & 0.01 \\
V & 0.09 & 0.05 & -0.02 & -0.03  & 0.12 & 0.07 & -0.05 & -0.09  & 0.10 & 0.04 & 0.02 & 0.01  & 0.00 & 0.00 & 0.00 & 0.00 \\
Cr & -0.12 & -0.06 & 0.09 & 0.17  & 0.17 & 0.09 & -0.07 & -0.12  & 0.03 & 0.00 & 0.06 & 0.07  & 0.00 & 0.00 & 0.00 & 0.00 \\
Mn & -0.08 & -0.05 & 0.06 & 0.14  & 0.16 & 0.08 & -0.07 & -0.13  & 0.04 & 0.01 & 0.05 & 0.08  & 0.00 & 0.00 & -0.01 & 0.00 \\
Ni & 0.35 & 0.08 & 0.21 & 0.23  & 0.31 & 0.23 & -0.06 & 0.01  & 0.23 & -0.05 & 0.21 & 0.26  & 0.00 & 0.00 & 0.01 & 0.01 \\
\hline
Assuming $T_{\rm eff}$ = 3600 K \\
Fe (FeI) & -0.12 & -0.06 & 0.08 & 0.17  & 0.18 & 0.08 & -0.06 & -0.13  & 0.05 & 0.02 & 0.06 & 0.10  & 0.00 & 0.00 & 0.00 & 0.00 \\
Fe (FeH) & 0.06 & 0.03 & -0.02 & -0.03  & 0.03 & 0.01 & -0.01 & -0.02  & 0.13 & 0.07 & -0.07 & -0.13  & -0.05 & -0.02 & 0.02 & 0.04 \\
C & 0.02 & 0.01 & -0.01 & -0.02  & 0.03 & 0.02 & -0.02 & -0.05  & 0.08 & -0.01 & -0.08 & -0.16  & 0.00 & 0.00 & 0.00 & -0.01 \\
O (H$_{2}$O) & 0.08 & 0.04 & -0.03 & -0.06  & -0.01 & 0.00 & 0.00 & 0.00  & 0.02 & 0.10 & -0.10 & -0.32  & 0.00 & -0.01 & -0.01 & 0.00 \\
O (OH) & -0.01 & 0.00 & 0.00 & 0.00  & 0.01 & 0.01 & -0.02 & -0.05  & 0.06 & 0.00 & -0.09 & -0.18  & -0.02 & -0.01 & 0.00 & 0.01 \\
Na & -0.02 & -0.01 & -0.01 & -0.04  & 0.08 & 0.02 & -0.01 & -0.02  & 0.20 & 0.09 & -0.01 & 0.00  & 0.00 & 0.00 & 0.00 & 0.00 \\
Mg & -0.21 & -0.11 & 0.11 & 0.20  & 0.13 & 0.07 & -0.08 & -0.16  & -0.04 & -0.03 & 0.02 & 0.02  & -0.01 & -0.01 & 0.00 & 0.00 \\
Al & -0.13 & -0.05 & 0.06 & 0.10  & 0.07 & 0.03 & -0.04 & -0.08  & -0.01 & 0.00 & 0.00 & -0.01  & 0.00 & 0.00 & 0.00 & 0.01 \\
Si & -0.27 & -0.13 & 0.16 & 0.27  & 0.21 & 0.11 & -0.11 & -0.20  & -0.02 & -0.01 & 0.04 & 0.05  & -0.02 & -0.01 & 0.01 & 0.02 \\
K & 0.06 & 0.03 & -0.03 & -0.04  & 0.00 & 0.00 & -0.01 & -0.02  & 0.00 & 0.00 & 0.00 & 0.01  & -0.01 & -0.01 & 0.00 & 0.00 \\
Ca & -0.05 & -0.02 & 0.02 & 0.04  & 0.10 & 0.05 & -0.05 & -0.11  & -0.06 & -0.03 & 0.04 & 0.08  & 0.00 & 0.00 & 0.00 & 0.00 \\
Ti & -0.03 & -0.01 & 0.02 & 0.04  & 0.04 & 0.02 & -0.02 & -0.04  & 0.07 & 0.03 & -0.03 & -0.07  & -0.01 & 0.00 & 0.01 & 0.03 \\
V & 0.10 & 0.05 & -0.02 & -0.03  & 0.13 & 0.08 & -0.05 & -0.09  & 0.11 & 0.06 & 0.04 & 0.02  & 0.00 & 0.00 & 0.00 & 0.00 \\
%Cr &  &  &  &   &  &  &  &   &  &  &  &   &  &  &  &  \\
Mn & 0.11 & 0.03 & 0.01 & 0.02  & 0.14 & 0.06 & -0.05 & -0.09  & 0.14 & 0.07 & 0.07 & 0.14  & 0.00 & 0.00 & 0.00 & 0.00 \\
%Ni &  &  &  &   &  &  &  &   &  &  &  &   &  &  &  &  \\
\hline
Assuming $T_{\rm eff}$ = 3500 K \\
Fe (FeI) & -0.18 & -0.09 & 0.09 & 0.19  & 0.09 & 0.05 & -0.07 & -0.12  & 0.00 & 0.00 & -0.01 & 0.00  & 0.00 & 0.00 & 0.00 & -0.01 \\
Fe (FeH) & 0.07 & 0.03 & -0.03 & -0.06  & 0.02 & 0.01 & -0.01 & -0.01  & 0.15 & 0.07 & -0.07 & -0.15  & -0.05 & -0.03 & 0.02 & 0.05 \\
C & 0.01 & 0.00 & -0.01 & -0.03  & 0.01 & 0.00 & -0.02 & -0.03  & 0.06 & -0.02 & -0.09 & -0.18  & 0.00 & 0.00 & -0.01 & -0.01 \\
O (H$_{2}$O) & 0.07 & 0.03 & -0.03 & -0.06  & 0.00 & 0.00 & -0.01 & -0.15  & 0.02 & -0.04 & -0.10 & -0.20  & -0.15 & -0.01 & 0.00 & 0.00 \\
O (OH) & -0.01 & 0.00 & 0.00 & 0.01  & 0.02 & 0.01 & -0.04 & -0.07  & 0.06 & 0.01 & -0.08 & -0.19  & -0.04 & -0.02 & 0.01 & 0.02 \\
Na & -0.03 & -0.01 & 0.03 & 0.06  & 0.05 & 0.02 & 0.00 & -0.01  & 0.19 & 0.09 & -0.06 & -0.08  & 0.01 & 0.01 & 0.00 & 0.00 \\
Mg & -0.20 & -0.11 & 0.10 & 0.20  & 0.14 & 0.07 & -0.08 & -0.15  & -0.03 & -0.02 & 0.01 & 0.02  & 0.00 & 0.00 & 0.00 & 0.00 \\
Al & -0.14 & -0.07 & 0.06 & 0.11  & 0.07 & 0.04 & -0.04 & -0.08  & 0.00 & 0.00 & 0.00 & -0.02  & 0.00 & 0.00 & 0.01 & 0.01 \\
Si & -0.29 & -0.15 & 0.19 & 0.30  & 0.27 & 0.14 & -0.10 & -0.20  & -0.05 & -0.02 & 0.06 & 0.09  & -0.01 & -0.01 & 0.02 & 0.03 \\
K & 0.07 & 0.04 & -0.03 & -0.06  & -0.02 & -0.01 & 0.00 & 0.00  & 0.01 & 0.00 & 0.00 & -0.01  & -0.01 & -0.01 & 0.00 & 0.00 \\
Ca & -0.07 & -0.03 & 0.02 & 0.04  & 0.11 & 0.05 & -0.05 & -0.11  & -0.06 & -0.03 & 0.03 & 0.07  & -0.01 & 0.00 & 0.00 & 0.00 \\
Ti & -0.03 & -0.01 & 0.02 & 0.03  & 0.03 & 0.02 & -0.02 & -0.04  & 0.07 & 0.03 & -0.05 & -0.10  & -0.01 & -0.01 & 0.01 & 0.02 \\
V & 0.10 & 0.05 & -0.04 & -0.06  & 0.16 & 0.08 & -0.07 & -0.12  & 0.12 & 0.06 & 0.02 & 0.04  & 0.00 & 0.00 & 0.00 & 0.00 \\
%Cr &  &  &  &   &  &  &  &   &  &  &  &   & 0.00 & 0.00 & 0.00 & 0.00 \\
Mn & 0.13 & 0.08 & -0.02 & -0.04  & 0.09 & 0.04 & -0.03 & -0.07  & 0.26 & 0.11 & 0.03 & 0.14  & 0.00 & 0.00 & 0.01 & 0.01 \\
%Ni &  &  &  &   &  &  &  &   &  &  &  &   &  &  &  &  \\
\hline
Assuming $T_{\rm eff}$ = 3400 K \\
Fe (FeI) & -0.18 & -0.09 & 0.10 & 0.19  & 0.11 & 0.05 & -0.06 & -0.11  & 0.00 & 0.00 & 0.01 & 0.02  & 0.00 & 0.00 & 0.00 & -0.01 \\
Fe (FeH) & 0.07 & 0.04 & -0.03 & -0.06  & 0.01 & 0.00 & 0.00 & 0.00  & 0.15 & 0.09 & -0.07 & -0.15  & -0.06 & -0.03 & 0.03 & 0.06 \\
C & 0.01 & 0.01 & 0.00 & -0.01  & 0.02 & 0.01 & 0.00 & -0.01  & 0.06 & -0.03 & -0.09 & -0.18  & 0.00 & 0.00 & 0.00 & 0.00 \\
O (H$_{2}$O) & 0.06 & 0.03 & -0.03 & -0.05  & 0.00 & 0.00 & -0.01 & -0.01  & 0.16 & 0.09 & -0.09 & -0.19  & 0.00 & 0.00 & 0.00 & 0.00 \\
O (OH) & -0.02 & -0.01 & 0.00 & 0.02  & 0.03 & 0.02 & -0.02 & -0.08  & 0.06 & -0.01 & -0.08 & -0.18  & -0.03 & -0.01 & 0.01 & 0.02 \\
Na & -0.01 & -0.03 & 0.03 & 0.05  & 0.04 & 0.01 & -0.07 & -0.08  & 0.04 & 0.00 & -0.07 & -0.11  & 0.01 & 0.01 & -0.04 & -0.03 \\
Mg & -0.17 & -0.09 & 0.10 & 0.21  & 0.15 & 0.07 & -0.07 & -0.14  & -0.01 & 0.00 & 0.01 & 0.02  & 0.00 & 0.00 & -0.01 & -0.01 \\
Al & -0.13 & -0.06 & 0.08 & 0.15  & 0.08 & 0.04 & -0.03 & -0.07  & 0.02 & 0.01 & -0.01 & -0.03  & 0.00 & 0.00 & 0.01 & 0.02 \\
Si & -0.21 & -0.18 & 0.20 & 0.33  & 0.33 & 0.18 & -0.15 & -0.24  & -0.10 & -0.06 & 0.09 & 0.15  & -0.03 & -0.02 & 0.01 & 0.03 \\
K & 0.10 & 0.06 & -0.03 & -0.07  & -0.02 & -0.01 & 0.02 & 0.03  & 0.03 & 0.02 & -0.01 & -0.02  & 0.00 & 0.00 & 0.01 & 0.01 \\
Ca & -0.10 & -0.04 & 0.04 & 0.06  & 0.13 & 0.06 & -0.06 & -0.13  & -0.06 & -0.03 & 0.03 & 0.06  & 0.00 & 0.00 & 0.00 & 0.00 \\
Ti & -0.03 & -0.01 & 0.02 & 0.04  & 0.03 & 0.02 & -0.01 & -0.03  & 0.07 & 0.04 & -0.06 & -0.12  & -0.01 & 0.00 & 0.01 & 0.02 \\
%V &  &  &  &   &  &  &  &   &  &  &  &   &  &  &  &  \\
%Cr &  &  &  &   &  &  &  &   &  &  &  &   &  &  &  &  \\
%Mn &  &  &  &   &  &  &  &   &  &  &  &   &  &  &  &  \\
%Ni &  &  &  &   &  &  &  &   &  &  &  &   &  &  &  &  \\
\hline
Assuming $T_{\rm eff}$ = 3300 K \\
Fe (FeI) &  &  &  &   &  &  &  &   &  &  &  &   &  &  &  &  \\
Fe (FeH) & 0.08 & 0.04 & -0.03 & -0.06  & 0.00 & 0.00 & 0.00 & 0.01  & 0.17 & 0.08 & -0.50 & -0.16  & -0.06 & -0.03 & 0.03 & 0.06 \\
C & 0.01 & 0.01 & 0.00 & 0.00  & 0.02 & 0.01 & 0.00 & 0.00  & 0.06 & -0.02 & -0.08 & -0.17  & 0.00 & 0.00 & 0.01 & 0.01 \\
O (H$_{2}$O) & 0.05 & 0.02 & -0.03 & -0.05  & 0.01 & 0.00 & 0.00 & -0.01  & 0.14 & 0.09 & -0.10 & -0.19  & 0.00 & 0.00 & 0.00 & 0.00 \\
O (OH) & -0.01 & -0.01 & 0.01 & 0.02  & 0.02 & 0.01 & -0.02 & -0.07  & 0.06 & 0.01 & -0.09 & -0.17  & -0.02 & -0.01 & 0.01 & 0.02 \\
Na & 0.02 & 0.01 & -0.02 & 0.08  & 0.06 & 0.02 & -0.02 & -0.04  & 0.08 & 0.04 & -0.03 & 0.19  & -0.02 & -0.01 & 0.00 & 0.01 \\
Mg & -0.15 & -0.08 & 0.10 & 0.22  & 0.16 & 0.08 & -0.06 & -0.12  & 0.02 & 0.01 & 0.02 & 0.03  & 0.00 & 0.00 & 0.00 & 0.00 \\
Al & -0.14 & -0.06 & 0.07 & 0.14  & 0.08 & 0.04 & -0.04 & -0.08  & 0.02 & 0.00 & -0.02 & -0.04  & 0.00 & 0.00 & 0.01 & 0.01 \\
Si & -0.08 & -0.09 & 0.35 & 0.51  & 0.63 & 0.35 & -0.09 & -0.10  & -0.03 & -0.05 & 0.25 & 0.43  & -0.01 & 0.00 & 0.00 & -0.01 \\
K & 0.11 & 0.05 & -0.04 & -0.10  & -0.06 & -0.03 & 0.02 & 0.04  & 0.03 & 0.02 & -0.03 & -0.05  & -0.01 & 0.00 & 0.00 & 0.00 \\
Ca & -0.12 & -0.06 & 0.05 & 0.09  & 0.14 & 0.07 & -0.08 & -0.15  & -0.05 & -0.03 & 0.02 & 0.06  & 0.00 & -0.01 & 0.00 & 0.00 \\
Ti & -0.04 & -0.01 & 0.02 & 0.04  & 0.03 & 0.01 & -0.01 & -0.03  & 0.08 & 0.04 & -0.07 & -0.15  & -0.01 & 0.00 & 0.01 & 0.02 \\
%V &  &  &  &   &  &  &  &   &  &  &  &   &  &  &  &  \\
%Cr &  &  &  &   &  &  &  &   &  &  &  &   &  &  &  &  \\
%Mn &  &  &  &   &  &  &  &   &  &  &  &   &  &  &  &  \\
%Ni &  &  &  &   &  &  &  &   &  &  &  &   &  &  &  &  \\
\hline
Assuming $T_{\rm eff}$ = 3200 K \\
Fe (FeI) &  &  &  &   &  &  &  &   &  &  &  &   &  &  &  &  \\
Fe (FeH) & 0.09 & 0.05 & -0.03 & -0.06  & 0.00 & 0.00 & 0.01 & 0.02  & 0.18 & 0.09 & -0.08 & -0.17  & -0.07 & -0.03 & 0.03 & 0.06 \\
C & 0.02 & 0.01 & 0.00 & 0.00  & 0.01 & 0.01 & 0.01 & 0.00  & 0.08 & 0.01 & -0.09 & -0.18  & 0.01 & 0.01 & 0.01 & 0.01 \\
O (H$_{2}$O) & 0.05 & 0.03 & -0.03 & -0.05  & 0.00 & 0.00 & 0.00 & 0.00  & 0.13 & 0.07 & -0.10 & -0.19  & 0.00 & 0.00 & 0.00 & 0.00 \\
O (OH) & 0.00 & 0.00 & 0.00 & -0.01  & 0.00 & 0.00 & -0.01 & -0.07  & 0.07 & 0.01 & -0.10 & -0.18  & -0.01 & -0.01 & 0.01 & 0.01 \\
Na & 0.03 & 0.01 & -0.02 & -0.03  & 0.04 & 0.01 & -0.02 & -0.04  & 0.54 & 0.05 & -0.04 & -0.02  & -0.02 & -0.01 & 0.00 & 0.00 \\
Mg & -0.14 & -0.07 & 0.08 & 0.18  & 0.16 & 0.07 & -0.07 & -0.13  & 0.03 & 0.01 & 0.00 & 0.03  & -0.01 & -0.01 & -0.01 & -0.01 \\
Al & -0.15 & -0.08 & 0.07 & 0.13  & 0.07 & 0.04 & -0.06 & -0.11  & 0.00 & 0.00 & -0.03 & -0.07  & -0.02 & -0.02 & 0.00 & 0.00 \\
Si & 0.75 & 0.00 & 0.52 & 0.75  & -0.06 & -0.03 & 0.04 & 0.53  & 0.10 & 0.04 & -0.15 & -0.29  & 0.02 & 0.01 & -0.02 & -0.04 \\
K & 0.13 & 0.06 & -0.07 & -0.12  & -0.08 & -0.04 & 0.03 & 0.05  & 0.03 & 0.02 & -0.05 & -0.10  & -0.02 & -0.01 & 0.00 & 0.00 \\
Ca & -0.16 & -0.08 & 0.06 & 0.11  & 0.15 & 0.07 & -0.10 & -0.17  & -0.06 & -0.03 & 0.01 & 0.04  & -0.02 & -0.02 & -0.01 & -0.02 \\
Ti & -0.02 & -0.01 & 0.02 & 0.04  & 0.01 & 0.01 & 0.00 & -0.02  & 0.08 & 0.04 & -0.08 & -0.18  & -0.01 & 0.00 & 0.00 & 0.01 \\
%V &  &  &  &   &  &  &  &   &  &  &  &   &  &  &  &  \\
%Cr &  &  &  &   &  &  &  &   &  &  &  &   &  &  &  &  \\
%Mn &  &  &  &   &  &  &  &   &  &  &  &   &  &  &  &  \\
%Ni &  &  &  &   &  &  &  &   &  &  &  &   &  &  &  &  \\
\hline
\tablewidth{0pt}	
\enddata
\tablenotetext{}{}
%\end{longtable}
\end{deluxetable*}
\end{longrotatetable}

%-------------------------------------
% \begin{longtable}
\startlongtable
\begin{deluxetable*}{lccccccccc}
%\rotate
\label{tab:noise} 
\tabletypesize{\scriptsize}
\tablecaption{Uncertanties due to signal-to-noise ratio (SNR) changes.}
\tablewidth{0pt}
\tablehead{
\colhead{SNR} &
\colhead{4000K} &
\colhead{3900K} &
\colhead{3800K} &
\colhead{3700K} &
\colhead{3600K} &
\colhead{3500K} &
\colhead{3400K} &
\colhead{3300K} &
\colhead{3200K} 
}
% \hline
\startdata
A(Fe) from Fe I lines\\
40 & 0.03 & 0.03 & 0.03 & 0.06 & 0.07 & 0.10 & 0.10 & 0.14 & 0.20  \\
60 & 0.02 & 0.02 & 0.02 & 0.04 & 0.06 & 0.09 & 0.07 & 0.09 & 0.08   \\
80 & 0.01 & 0.01 & 0.02 & 0.02 & 0.04 & 0.08 & 0.04 & 0.06 & 0.06  \\
100 & 0.01 & 0.01 & 0.01 & 0.02 & 0.04 & 0.07 & 0.04 & 0.04 & 0.05  \\
120 & 0.01 & 0.01 & 0.01 & 0.02 & 0.05 & 0.05 & 0.03 & 0.07 & 0.05  \\
140 & 0.01 & 0.01 & 0.01 & 0.02 & 0.05 & 0.05 & 0.03 & 0.05 & 0.04  \\
160 & 0.01 & 0.01 & 0.01 & 0.01 & 0.03 & 0.05 & 0.03 & 0.05 & 0.04  \\
180 & 0.01 & 0.01 & 0.01 & 0.01 & 0.03 & 0.04 & 0.02 & 0.04 & 0.04  \\
200 & 0.01 & 0.01 & 0.01 & 0.01 & 0.04 & 0.03 & 0.02 & 0.03 & 0.04  \\
220 & 0.00 & 0.01 & 0.01 & 0.01 & 0.04 & 0.04 & 0.03 & 0.03 & 0.03  \\
240 & 0.00 & 0.01 & 0.01 & 0.01 & 0.01 & 0.02 & 0.02 & 0.03 & 0.03  \\
260 & 0.00 & 0.00 & 0.01 & 0.01 & 0.02 & 0.01 & 0.02 & 0.03 & 0.03  \\
280 & 0.00 & 0.01 & 0.01 & 0.01 & 0.02 & 0.02 & 0.02 & 0.03 & 0.03  \\
300+ & 0.00 & 0.00 & 0.00 & 0.01 & 0.02 & 0.01 & 0.02 & 0.03 & 0.02  \\
\hline
A(Fe) from FeH lines\\
40 & 0.03 & 0.03 & 0.04 & 0.04 & 0.04 & 0.03 & 0.03 & 0.03 & 0.03  \\
60 & 0.03 & 0.02 & 0.03 & 0.02 & 0.03 & 0.03 & 0.02 & 0.02 & 0.02  \\
80 & 0.02 & 0.02 & 0.02 & 0.01 & 0.02 & 0.02 & 0.02 & 0.02 & 0.02  \\
100 & 0.02 & 0.02 & 0.01 & 0.01 & 0.02 & 0.02 & 0.02 & 0.02 & 0.01  \\
120 & 0.01 & 0.02 & 0.02 & 0.01 & 0.01 & 0.01 & 0.01 & 0.01 & 0.01  \\
140 & 0.01 & 0.01 & 0.01 & 0.01 & 0.01 & 0.01 & 0.01 & 0.01 & 0.01  \\
160 & 0.01 & 0.01 & 0.01 & 0.01 & 0.01 & 0.01 & 0.01 & 0.01 & 0.01  \\
180 & 0.01 & 0.01 & 0.01 & 0.01 & 0.01 & 0.01 & 0.01 & 0.01 & 0.01  \\
200 & 0.01 & 0.01 & 0.01 & 0.01 & 0.01 & 0.01 & 0.01 & 0.01 & 0.01  \\
220 & 0.01 & 0.01 & 0.01 & 0.01 & 0.00 & 0.01 & 0.01 & 0.01 & 0.01  \\
240 & 0.01 & 0.01 & 0.01 & 0.01 & 0.01 & 0.00 & 0.00 & 0.01 & 0.01  \\
260 & 0.01 & 0.01 & 0.01 & 0.01 & 0.01 & 0.00 & 0.01 & 0.01 & 0.01  \\
280 & 0.01 & 0.01 & 0.01 & 0.01 & 0.01 & 0.01 & 0.01 & 0.01 & 0.01  \\
300+ & 0.01 & 0.01 & 0.00 & 0.01 & 0.01 & 0.00 & 0.01 & 0.00 & 0.01  \\
\hline
A(O) from OH lines\\
40 & 0.01 & 0.01 & 0.01 & 0.01 & 0.02 & 0.02 & 0.01 & 0.02 & 0.04  \\
60 & 0.01 & 0.01 & 0.01 & 0.01 & 0.01 &  0.01 & 0.02 & 0.02 & 0.02  \\
80 & 0.01 & 0.01 & 0.01 & 0.01 & 0.01 & 0.01 & 0.01 & 0.01 & 0.01  \\
100 & 0.01 & 0.01 & 0.01 & 0.01 & 0.01 & 0.01 & 0.01 & 0.01 & 0.01  \\
120 & 0.01 & 0.00 & 0.00 & 0.00 & 0.00 & 0.01 & 0.01 & 0.01 & 0.01  \\
140 & 0.00 & 0.00 & 0.00 & 0.00 & 0.00 & 0.01 & 0.01 & 0.01 & 0.01  \\
160 & 0.00 & 0.00 & 0.00 & 0.00 & 0.00 & 0.00 & 0.00 & 0.01 & 0.01  \\
180 & 0.00 & 0.00 & 0.00 & 0.00 & 0.00 & 0.00 & 0.00 & 0.01 & 0.01  \\
200 & 0.00 & 0.00 & 0.00 & 0.00 & 0.00 & 0.00 & 0.00 & 0.01 & 0.01  \\
220 & 0.00 & 0.00 & 0.00 & 0.00 & 0.00 & 0.00 & 0.00 & 0.00 & 0.01  \\
240 & 0.00 & 0.00 & 0.00 & 0.00 & 0.00 & 0.00 & 0.00 & 0.00 & 0.01  \\
260 & 0.00 & 0.00 & 0.00 & 0.00 & 0.00 & 0.00 & 0.00 & 0.01 & 0.00  \\
280 & 0.00 & 0.00 & 0.00 & 0.00 & 0.00 & 0.00 & 0.00 & 0.00 & 0.00  \\
300+ & 0.00 & 0.00 & 0.00 & 0.00 & 0.00 & 0.00 & 0.00 & 0.00 & 0.00  \\
\hline
A(O) from H$_{2}$O lines\\
40 & 0.02 & 0.02 & 0.06 & 0.06 & 0.03 & 0.03 & 0.02 & 0.02 & 0.03  \\
60 & 0.01 & 0.01 & 0.05 & 0.03 & 0.02 & 0.02 & 0.02 & 0.02 & 0.02  \\
80 & 0.01 & 0.01 & 0.03 & 0.03 & 0.02 & 0.01 & 0.01 & 0.02 & 0.01  \\
100 & 0.01 & 0.01 & 0.03 & 0.02 & 0.02 & 0.01 & 0.01 & 0.01 & 0.01  \\
120 & 0.01 & 0.01 & 0.04 & 0.02 & 0.01 & 0.01 & 0.01 & 0.01 & 0.01  \\
140 & 0.01 & 0.01 & 0.02 & 0.02 & 0.01 & 0.01 & 0.01 & 0.01 & 0.01  \\
160 & 0.01 & 0.01 & 0.02 & 0.01 & 0.01 & 0.01 & 0.01 & 0.01 & 0.01  \\
180 & 0.00 & 0.00 & 0.02 & 0.01 & 0.01 & 0.01 & 0.00 & 0.01 & 0.01  \\
200 & 0.00 & 0.00 & 0.02 & 0.01 & 0.01 & 0.01 & 0.00 & 0.01 & 0.01  \\
220 & 0.00 & 0.00 & 0.01 & 0.01 & 0.01 & 0.00 & 0.00 & 0.01 & 0.00  \\
240 & 0.00 & 0.00 & 0.01 & 0.01 & 0.01 & 0.00 & 0.00 & 0.01 & 0.00  \\
260 & 0.00 & 0.00 & 0.01 & 0.01 & 0.01 & 0.00 & 0.01 & 0.00 & 0.00  \\
280 & 0.00 & 0.00 & 0.01 & 0.01 & 0.01 & 0.00 & 0.00 & 0.00 & 0.00  \\
300+ & 0.00 & 0.00 & 0.01 & 0.01 & 0.00 & 0.00 & 0.00 & 0.00 & 0.00  \\
\hline
A(C) from CO lines\\
40 & 0.09 & 0.08 & 0.10 & 0.05 & 0.05 & 0.20 & 0.22 & 0.16 & 0.20  \\
60 & 0.07 & 0.07 & 0.08 & 0.05 & 0.05 & 0.11 & 0.24 & 0.12 & 0.24  \\
80 & 0.06 & 0.07 & 0.06 & 0.03 & 0.03 & 0.10 & 0.15 & 0.10 & 0.13  \\
100 & 0.05 & 0.06 & 0.04 & 0.03 & 0.03 & 0.10 & 0.13 & 0.07 & 0.12  \\
120 & 0.04 & 0.04 & 0.05 & 0.02 & 0.02 & 0.08 & 0.07 & 0.06 & 0.11  \\
140 & 0.05 & 0.04 & 0.05 & 0.01 & 0.01 & 0.06 & 0.05 & 0.04 & 0.06  \\
160 & 0.04 & 0.04 & 0.03 & 0.01 & 0.01 & 0.04 & 0.04 & 0.05 & 0.05  \\
180 & 0.04 & 0.04 & 0.04 & 0.01 & 0.01 & 0.03 & 0.04 & 0.03 & 0.04  \\
200 & 0.03 & 0.03 & 0.04 & 0.02 & 0.02 & 0.05 & 0.03 & 0.04 & 0.04  \\
220 & 0.03 & 0.03 & 0.02 & 0.01 & 0.01 & 0.05 & 0.03 & 0.03 & 0.04  \\
240 & 0.02 & 0.03 & 0.02 & 0.01 & 0.01 & 0.03 & 0.03 & 0.05 & 0.04  \\
260 & 0.03 & 0.03 & 0.02 & 0.01 & 0.01 & 0.03 & 0.03 & 0.03 & 0.04  \\
280 & 0.03 & 0.03 & 0.03 & 0.01 & 0.01 & 0.03 & 0.03 & 0.03 & 0.04  \\
300+ & 0.02 & 0.03 & 0.02 & 0.01 & 0.01 & 0.03 & 0.03 & 0.04 & 0.03  \\
\hline
A(Na) from Na I lines\\
40 & 0.16 & 0.13 & 0.23 & 0.26 & 0.22 & 0.10 & 0.15 & 0.84 & 0.28  \\
60 & 0.09 & 0.12 & 0.08 & 0.19 & 0.12 & 0.08 & 0.13 & 0.26 & 0.20  \\
80 & 0.07 & 0.11 & 0.09 & 0.13 & 0.11 & 0.09 & 0.11 & 0.16 & 0.14  \\
100 & 0.07 & 0.06 & 0.06 & 0.12 & 0.07 & 0.05 & 0.03 & 0.14 & 0.12  \\
120 & 0.02 & 0.06 & 0.06 & 0.10 & 0.06 & 0.05 & 0.03 & 0.12 & 0.11  \\
140 & 0.02 & 0.08 & 0.02 & 0.06 & 0.06 & 0.03 & 0.02 & 0.12 & 0.08  \\
160 & 0.01 & 0.02 & 0.05 & 0.07 & 0.05 & 0.04 & 0.03 & 0.11 & 0.04  \\
180 & 0.01 & 0.02 & 0.02 & 0.07 & 0.04 & 0.03 & 0.03 & 0.09 & 0.06  \\
200 & 0.02 & 0.01 & 0.02 & 0.03 & 0.05 & 0.02 & 0.02 & 0.09 & 0.04  \\
220 & 0.01 & 0.02 & 0.01 & 0.04 & 0.03 & 0.02 & 0.02 & 0.09 & 0.01  \\
240 & 0.01 & 0.01 & 0.01 & 0.05 & 0.03 & 0.02 & 0.02 & 0.08 & 0.01  \\
260 & 0.01 & 0.01 & 0.01 & 0.01 & 0.03 & 0.02 & 0.01 & 0.01 & 0.01  \\
280 & 0.01 & 0.01 & 0.01 & 0.01 & 0.02 & 0.02 & 0.01 & 0.01 & 0.01  \\
300+ & 0.01 & 0.01 & 0.01 & 0.03 & 0.03 & 0.02 & 0.01 & 0.01 & 0.01  \\
\hline
A(Mg) from Mg I lines\\
40 & 0.04 & 0.06 & 0.05 & 0.06 & 0.07 & 0.06 & 0.07 & 0.10 & 0.14  \\
60 & 0.04 & 0.03 & 0.05 & 0.04 & 0.04 & 0.04 & 0.05 & 0.09 & 0.08  \\
80 & 0.03 & 0.03 & 0.02 & 0.02 & 0.03 & 0.03 & 0.04 & 0.08 & 0.05  \\
100 & 0.02 & 0.02 & 0.03 & 0.01 & 0.02 & 0.02 & 0.04 & 0.04 & 0.05  \\
120 & 0.02 & 0.02 & 0.02 & 0.02 & 0.02 & 0.02 & 0.03 & 0.04 & 0.04  \\
140 & 0.01 & 0.02 & 0.02 & 0.02 & 0.02 & 0.02 & 0.02 & 0.03 & 0.03  \\
160 & 0.01 & 0.02 & 0.01 & 0.02 & 0.01 & 0.02 & 0.02 & 0.03 & 0.03  \\
180 & 0.01 & 0.01 & 0.02 & 0.01 & 0.01 & 0.02 & 0.02 & 0.02 & 0.03  \\
200 & 0.01 & 0.01 & 0.01 & 0.01 & 0.01 & 0.02 & 0.02 & 0.02 & 0.03  \\
220 & 0.01 & 0.01 & 0.01 & 0.01 & 0.01 & 0.01 & 0.01 & 0.02 & 0.02  \\
240 & 0.01 & 0.01 & 0.01 & 0.01 & 0.01 & 0.01 & 0.02 & 0.02 & 0.02  \\
260 & 0.01 & 0.01 & 0.01 & 0.01 & 0.01 & 0.01 & 0.01 & 0.01 & 0.02  \\
280 & 0.01 & 0.01 & 0.01 & 0.01 & 0.01 & 0.01 & 0.01 & 0.01 & 0.01  \\
300+ & 0.01 & 0.01 & 0.01 & 0.01 & 0.01 & 0.01 & 0.01 & 0.01 & 0.02  \\
\hline
A(Al) from Al I lines\\
40 & 0.04 & 0.03 & 0.04 & 0.05 & 0.05 & 0.05 & 0.05 & 0.06 & 0.06  \\
60 & 0.02 & 0.02 & 0.03 & 0.03 & 0.05 & 0.03 & 0.04 & 0.03 & 0.04  \\
80 & 0.02 & 0.02 & 0.02 & 0.02 & 0.02 & 0.03 & 0.03 & 0.04 & 0.03  \\
100 & 0.02 & 0.01 & 0.02 & 0.02 & 0.02 & 0.02 & 0.03 & 0.03 & 0.02  \\
120 & 0.01 & 0.01 & 0.02 & 0.02 & 0.02 & 0.02 & 0.02 & 0.02 & 0.01  \\
140 & 0.01 & 0.01 & 0.01 & 0.02 & 0.01 & 0.01 & 0.02 & 0.01 & 0.02  \\
160 & 0.01 & 0.01 & 0.01 & 0.01 & 0.01 & 0.01 & 0.02 & 0.01 & 0.02  \\
180 & 0.01 & 0.01 & 0.01 & 0.01 & 0.01 & 0.01 & 0.01 & 0.01 & 0.01  \\
200 & 0.01 & 0.01 & 0.01 & 0.01 & 0.01 & 0.01 & 0.01 & 0.01 & 0.01  \\
220 & 0.01 & 0.01 & 0.01 & 0.01 & 0.01 & 0.01 & 0.01 & 0.01 & 0.01  \\
240 & 0.01 & 0.01 & 0.01 & 0.01 & 0.01 & 0.01 & 0.01 & 0.01 & 0.01  \\
260 & 0.01 & 0.01 & 0.01 & 0.01 & 0.01 & 0.01 & 0.01 & 0.01 & 0.01  \\
280 & 0.01 & 0.01 & 0.01 & 0.01 & 0.01 & 0.01 & 0.01 & 0.01 & 0.01  \\
300+ & 0.01 & 0.01 & 0.01 & 0.01 & 0.01 & 0.01 & 0.01 & 0.01 & 0.01  \\
\hline
A(Si) from Si I lines\\
40 & 0.04 & 0.06 & 0.05 & 0.07 & 0.10 & 0.19 & 0.20 & 0.32 & 0.91  \\
60 & 0.02 & 0.03 & 0.03 & 0.07 & 0.08 & 0.12 & 0.11 & 0.23 & 0.22  \\
80 & 0.02 & 0.03 & 0.03 & 0.04 & 0.07 & 0.07 & 0.13 & 0.26 & 0.23  \\
100 & 0.02 & 0.02 & 0.02 & 0.03 & 0.07 & 0.06 & 0.10 & 0.16 & 0.20  \\
120 & 0.01 & 0.01 & 0.02 & 0.02 & 0.03 & 0.05 & 0.09 & 0.15 & 0.18  \\
140 & 0.01 & 0.01 & 0.02 & 0.03 & 0.03 & 0.07 & 0.07 & 0.15 & 0.16  \\
160 & 0.01 & 0.01 & 0.02 & 0.03 & 0.05 & 0.04 & 0.06 & 0.15 & 0.16  \\
180 & 0.01 & 0.01 & 0.02 & 0.02 & 0.04 & 0.05 & 0.05 & 0.14 & 0.16  \\
200 & 0.01 & 0.01 & 0.01 & 0.02 & 0.03 & 0.04 & 0.07 & 0.13 & 0.13  \\
220 & 0.01 & 0.01 & 0.02 & 0.02 & 0.03 & 0.03 & 0.06 & 0.13 & 0.13  \\
240 & 0.01 & 0.01 & 0.02 & 0.01 & 0.03 & 0.03 & 0.05 & 0.12 & 0.12  \\
260 & 0.01 & 0.01 & 0.01 & 0.01 & 0.03 & 0.03 & 0.05 & 0.13 & 0.12  \\
280 & 0.01 & 0.01 & 0.01 & 0.01 & 0.03 & 0.03 & 0.04 & 0.13 & 0.12  \\
300+ & 0.01 & 0.01 & 0.01 & 0.01 & 0.02 & 0.03 & 0.05 & 0.11 & 0.12  \\
\hline
A(K) from K I lines\\
40 & 0.04 & 0.04 & 0.04 & 0.04 & 0.04 & 0.05 & 0.05 & 0.04 & 0.05  \\
60 & 0.03 & 0.03 & 0.03 & 0.03 & 0.04 & 0.04 & 0.04 & 0.02 & 0.02  \\
80 & 0.02 & 0.02 & 0.02 & 0.02 & 0.02 & 0.03 & 0.03 & 0.02 & 0.02  \\
100 & 0.02 & 0.02 & 0.02 & 0.02 & 0.02 & 0.02 & 0.02 & 0.03 & 0.02  \\
120 & 0.02 & 0.01 & 0.01 & 0.02 & 0.01 & 0.02 & 0.02 & 0.02 & 0.02  \\
140 & 0.01 & 0.01 & 0.01 & 0.01 & 0.01 & 0.01 & 0.01 & 0.02 & 0.02  \\
160 & 0.01 & 0.01 & 0.01 & 0.01 & 0.02 & 0.01 & 0.01 & 0.01 & 0.01  \\
180 & 0.01 & 0.01 & 0.01 & 0.01 & 0.01 & 0.01 & 0.01 & 0.01 & 0.01  \\
200 & 0.01 & 0.01 & 0.01 & 0.01 & 0.01 & 0.01 & 0.01 & 0.01 & 0.01  \\
220 & 0.01 & 0.01 & 0.01 & 0.01 & 0.01 & 0.01 & 0.01 & 0.01 & 0.01  \\
240 & 0.01 & 0.01 & 0.01 & 0.01 & 0.01 & 0.01 & 0.01 & 0.01 & 0.01  \\
260 & 0.01 & 0.01 & 0.01 & 0.01 & 0.01 & 0.01 & 0.01 & 0.01 & 0.01  \\
280 & 0.01 & 0.01 & 0.01 & 0.01 & 0.01 & 0.01 & 0.01 & 0.01 & 0.01  \\
300+ & 0.01 & 0.01 & 0.01 & 0.01 & 0.01 & 0.01 & 0.01 & 0.01 & 0.01  \\
\hline
A(Ca) from Ca I lines\\
40 & 0.03 & 0.04 & 0.04 & 0.05 & 0.04 & 0.05 & 0.06 & 0.07 & 0.09  \\
60 & 0.02 & 0.03 & 0.04 & 0.03 & 0.02 & 0.04 & 0.05 & 0.04 & 0.07  \\
80 & 0.01 & 0.03 & 0.02 & 0.02 & 0.02 & 0.02 & 0.03 & 0.04 & 0.03  \\
100 & 0.01 & 0.02 & 0.01 & 0.02 & 0.02 & 0.02 & 0.03 & 0.02 & 0.04  \\
120 & 0.01 & 0.01 & 0.01 & 0.01 & 0.02 & 0.02 & 0.02 & 0.03 & 0.02  \\
140 & 0.01 & 0.01 & 0.01 & 0.01 & 0.01 & 0.01 & 0.02 & 0.02 & 0.02  \\
160 & 0.01 & 0.01 & 0.01 & 0.01 & 0.01 & 0.01 & 0.02 & 0.02 & 0.02  \\
180 & 0.01 & 0.01 & 0.01 & 0.01 & 0.01 & 0.01 & 0.02 & 0.02 & 0.02  \\
200 & 0.01 & 0.01 & 0.01 & 0.01 & 0.01 & 0.01 & 0.01 & 0.02 & 0.02  \\
220 & 0.01 & 0.01 & 0.01 & 0.01 & 0.01 & 0.01 & 0.01 & 0.02 & 0.02  \\
240 & 0.01 & 0.01 & 0.01 & 0.01 & 0.01 & 0.01 & 0.01 & 0.01 & 0.01  \\
260 & 0.01 & 0.01 & 0.01 & 0.01 & 0.01 & 0.01 & 0.01 & 0.01 & 0.01  \\
280 & 0.01 & 0.01 & 0.01 & 0.01 & 0.01 & 0.01 & 0.01 & 0.01 & 0.01  \\
300+ & 0.01 & 0.00 & 0.01 & 0.01 & 0.01 & 0.01 & 0.01 & 0.01 & 0.01  \\
\hline
A(Ti) from Ti I lines\\
40 & 0.03 & 0.04 & 0.04 & 0.04 & 0.05 & 0.03 & 0.05 & 0.05 & 0.05  \\
60 & 0.03 & 0.03 & 0.03 & 0.03 & 0.03 & 0.03 & 0.04 & 0.03 & 0.04  \\
80 & 0.02 & 0.02 & 0.02 & 0.02 & 0.02 & 0.02 & 0.02 & 0.03 & 0.04  \\
100 & 0.01 & 0.02 & 0.02 & 0.02 & 0.02 & 0.02 & 0.02 & 0.02 & 0.03  \\
120 & 0.01 & 0.01 & 0.02 & 0.02 & 0.02 & 0.02 & 0.02 & 0.01 & 0.02  \\
140 & 0.01 & 0.01 & 0.01 & 0.02 & 0.01 & 0.01 & 0.02 & 0.02 & 0.02  \\
160 & 0.01 & 0.01 & 0.01 & 0.01 & 0.01 & 0.01 & 0.01 & 0.01 & 0.02  \\
180 & 0.01 & 0.01 & 0.01 & 0.01 & 0.01 & 0.01 & 0.01 & 0.01 & 0.02  \\
200 & 0.01 & 0.01 & 0.01 & 0.01 & 0.01 & 0.01 & 0.01 & 0.01 & 0.01  \\
220 & 0.01 & 0.01 & 0.01 & 0.01 & 0.01 & 0.01 & 0.01 & 0.01 & 0.01  \\
240 & 0.01 & 0.01 & 0.01 & 0.01 & 0.01 & 0.01 & 0.01 & 0.01 & 0.02  \\
260 & 0.01 & 0.00 & 0.01 & 0.01 & 0.01 & 0.01 & 0.01 & 0.01 & 0.01  \\
280 & 0.01 & 0.01 & 0.01 & 0.01 & 0.01 & 0.01 & 0.01 & 0.01 & 0.01  \\
300+ & 0.01 & 0.01 & 0.01 & 0.01 & 0.01 & 0.01 & 0.01 & 0.01 & 0.01  \\
\hline
A(V) from V I lines\\
40 & 0.15 & 0.14 & 0.12 & 0.11 & 0.11 & 0.22 &  & &   \\
60 & 0.09 & 0.09 & 0.10 & 0.11 & 0.11 & 0.18 & & &   \\
80 & 0.06 & 0.07 & 0.06 & 0.06 & 0.06 & 0.10 & & &   \\
100 & 0.06 & 0.07 & 0.04 & 0.09 & 0.09 & 0.08 & & &   \\
120 & 0.05 & 0.07 & 0.04 & 0.05 & 0.05 & 0.08 & & &   \\
140 & 0.04 & 0.03 & 0.03 & 0.06 & 0.06 & 0.07 & & &   \\
160 & 0.04 & 0.04 & 0.04 & 0.04 & 0.04 & 0.06 & & &   \\
180 & 0.03 & 0.04 & 0.03 & 0.03 & 0.03 & 0.05 & & &   \\
200 & 0.03 & 0.03 & 0.03 & 0.03 & 0.03 & 0.05 & & &   \\
220 & 0.02 & 0.02 & 0.02 & 0.03 & 0.03 & 0.05 & & &   \\
240 & 0.03 & 0.03 & 0.01 & 0.03 & 0.03 & 0.06 & & &   \\
260 & 0.02 & 0.02 & 0.02 & 0.02 & 0.02 & 0.05 & & &   \\
280 & 0.02 & 0.02 & 0.01 & 0.03 & 0.03 & 0.04 & & &   \\
300+ & 0.02 & 0.03 & 0.02 & 0.03 & 0.03 & 0.03 & & &   \\
\hline
A(Cr) from Cr I lines\\
40 & 0.11 & 0.14 & 0.07 & 0.20 & 0.20 & & & &   \\
60 & 0.07 & 0.08 & 0.07 & 0.10 & 0.10 & & & &   \\
80 & 0.05 & 0.05 & 0.07 & 0.07 & 0.07 & & & &   \\
100 & 0.05 & 0.04 & 0.06 & 0.07 & 0.07 & & & &   \\
120 & 0.03 & 0.02 & 0.04 & 0.06 & 0.06 & & & &   \\
140 & 0.03 & 0.04 & 0.04 & 0.06 & 0.06 & & & &   \\
160 & 0.03 & 0.03 & 0.03 & 0.04 & 0.04 & & & &   \\
180 & 0.03 & 0.03 & 0.04 & 0.04 & 0.04 & & & &   \\
200 & 0.02 & 0.03 & 0.02 & 0.04 & 0.04 & & & &   \\
220 & 0.02 & 0.04 & 0.03 & 0.03 & 0.03 & & & &   \\
240 & 0.02 & 0.02 & 0.03 & 0.03 & 0.03 & & & &   \\
260 & 0.02 & 0.01 & 0.02 & 0.03 & 0.03 & & & &   \\
280 & 0.02 & 0.02 & 0.02 & 0.02 & 0.02 & & & &   \\
300+ & 0.01 & 0.02 & 0.02 & 0.03 & 0.03 & & & &   \\
\hline
A(Mn) from Mn I lines\\
40 & 0.04 & 0.05 & 0.05 & 0.11 & 0.11 & & & &   \\
60 & 0.04 & 0.03 & 0.04 & 0.05 & 0.05 & & & &   \\
80 & 0.02 & 0.03 & 0.03 & 0.03 & 0.03 & & & &   \\
100 & 0.02 & 0.02 & 0.02 & 0.03 & 0.03 & & & &   \\
120 & 0.02 & 0.02 & 0.03 & 0.03 & 0.03 & & & &   \\
140 & 0.01 & 0.02 & 0.02 & 0.02 & 0.02 & & & &   \\
160 & 0.01 & 0.01 & 0.01 & 0.02 & 0.02 & & & &   \\
180 & 0.01 & 0.01 & 0.02 & 0.02 & 0.02 & & & &   \\
200 & 0.01 & 0.01 & 0.01 & 0.02 & 0.02 & & & &   \\
220 & 0.01 & 0.01 & 0.01 & 0.01 & 0.01 & & & &   \\
240 & 0.01 & 0.01 & 0.01 & 0.01 & 0.01 & & & &   \\
260 & 0.01 & 0.01 & 0.01 & 0.01 & 0.01 & & & &   \\
280 & 0.01 & 0.01 & 0.01 & 0.01 & 0.01 & & & &   \\
300+ & 0.01 & 0.01 & 0.01 & 0.01 & 0.01 & & & &   \\
\hline
A(Ni) from Ni I lines\\
40 & 0.12 & 0.14 & 0.15 & 0.22 & & & & &   \\
60 & 0.08 & 0.08 & 0.11 & 0.14 & & & & &   \\
80 & 0.07 & 0.10 & 0.11 & 0.13 & & & & &   \\
100 & 0.05 & 0.11 & 0.11 & 0.11 & & & & &   \\
120 & 0.04 & 0.06 & 0.08 & 0.13 & & & & &   \\
140 & 0.04 & 0.06 & 0.08 & 0.07 & & & & &   \\
160 & 0.03 & 0.06 & 0.06 & 0.07 & & & & &   \\
180 & 0.03 & 0.05 & 0.05 & 0.11 & & & & &   \\
200 & 0.02 & 0.05 & 0.07 & 0.06 & & & & &   \\
220 & 0.02 & 0.05 & 0.05 & 0.06 & & & & &   \\
240 & 0.02 & 0.04 & 0.05 & 0.05 & & & & &   \\
260 & 0.02 & 0.04 & 0.04 & 0.05 & & & & &   \\
280 & 0.02 & 0.03 & 0.05 & 0.04 & & & & &   \\
300+ & 0.01 & 0.03 & 0.05 & 0.04 & & & & &   \\
\hline
\tablewidth{0pt}	
\enddata
\tablenotetext{}{300+ indicates that SNR higher than 300 can also adopt the values.}
\end{deluxetable*}
% \end{longtable}

%-------------------------------------
% \begin{longtable}
\startlongtable
\begin{deluxetable*}{lccccccccc}
%\rotate
\label{tab:displacement}
\tabletypesize{\scriptsize}
\tablecaption{Uncertanties due to pseudo-continuum displacements.}
\tablewidth{0pt}
\tablehead{
\colhead{Pseudocontinuum displacement} &
\colhead{4000K} &
\colhead{3900K} &
\colhead{3800K} &
\colhead{3700K} &
\colhead{3600K} &
\colhead{3500K} &
\colhead{3400K} &
\colhead{3300K} &
\colhead{3200K} 
}
% \hline
\startdata
A(Fe) from Fe I lines\\
-2\% & -0.147 & -0.181 & -0.209 & -0.335 & -0.287 & -0.334 & -0.227 & -0.316 & -0.316   \\
-1\% & -0.072 & -0.088 & -0.098 & -0.160 & -0.135 & -0.163 & -0.113 & -0.155 & -0.155   \\
+1\% & 0.072 & 0.084 & 0.088 & 0.119 & 0.119 & 0.136 & 0.111 & 0.085 & 0.085  \\
+2\% & 0.137 & 0.157 & 0.168 & 0.209 & 0.206 & 0.226 & 0.211 & 0.156 & 0.156  \\
\hline
A(Fe) from FeH lines\\
-2\% & -0.224 & -0.226 & -0.227 & -0.252 & -0.226 & -0.190 & -0.196 & -0.202 & -0.177  \\
-1\% & -0.124 & -0.119 & -0.118 & -0.132 & -0.115 & -0.112 & -0.107 & -0.098 & -0.099  \\
+1\% & 0.100 & 0.099 & 0.098 & 0.101 & 0.079 & 0.083 & 0.083 & 0.099 & 0.078  \\
+2\% & 0.178 & 0.177 & 0.176 & 0.180 & 0.137 & 0.144 & 0.143 & 0.175 & 0.158  \\
\hline
A(O) from OH lines\\
-2\% & -0.033 & -0.031 & -0.027 & -0.033 & -0.040 & -0.057 & -0.056 & -0.055 & -0.065  \\
-1\% & -0.015 & -0.015 & -0.012 & -0.016 & -0.023 & -0.030 & -0.024 & -0.024 & -0.029  \\
+1\% & 0.027 & 0.026 & 0.028 & 0.037 & 0.036 & 0.022 & 0.017 & 0.030 & 0.025  \\
+2\% & 0.060 & 0.061 & 0.072 & 0.081 & 0.075 & 0.062 & 0.059 & 0.068 & 0.056  \\
\hline
A(O) from H$_{2}$O lines\\
-2\% &  & -0.087 & -0.087 & -0.073 & -0.068 & -0.049 & -0.049 & -0.039 & -0.039   \\
-1\% &  & -0.020 & -0.020 & -0.019 & -0.016 & -0.021 & -0.021 & -0.013 & -0.013   \\
+1\% &  & 0.037 & 0.037 & 0.019 & 0.014 & 0.013 & 0.013 & 0.021 & 0.021  \\
+2\% &  & 0.100 & 0.100 & 0.063 & 0.048 & 0.043 & 0.043 & 0.056 & 0.056  \\
\hline
A(C) from CO lines\\
-2\% & 0.024 & -0.107 & -0.363 & -0.123 & -0.093 & -0.099 & -0.082 &-0.070 & -0.011   \\
-1\% & -0.001 & -0.131 & -0.132 & -0.059 & -0.047 & -0.043 & -0.031 & -0.024 & -0.007   \\
+1\% & 0.018 & 0.017 & 0.018 & 0.046 & 0.046 & 0.047 & 0.034 & 0.023 & 0.009  \\
+2\% & 0.034 & 0.033 & 0.035 & 0.075 & 0.079 & 0.096 & 0.084 & 0.062 & 0.023  \\
\hline
A(Na) from Na I lines\\
-2\% & -0.455 & -0.428 & -0.417 & -0.199 & -0.199 & -0.199 & -0.417 & -0.111 & -0.067   \\
-1\% & -0.223 & -0.170 & -0.172 & -0.138 & -0.138 & -0.138 & -0.172 & -0.073 & -0.039   \\
+1\% & 0.061 & 0.064 & 0.063 & 0.099 & 0.099 & 0.099 & 0.063 & 0.145 & 0.064  \\
+2\% & 0.111 & 0.116 & 0.117 & 0.139 & 0.139 & 0.139 & 0.117 & 0.295 & 0.162   \\
\hline
A(Mg) from Mg I lines\\
-2\% & -0.108 & -0.088 & -0.113 & -0.201 & -0.172 & -0.184 & -0.246 & -0.224 & -0.328   \\
-1\% & -0.028 & -0.026 & -0.021 & -0.041 & -0.084 & -0.090 & -0.103 & -0.114 & -0.139  \\
+1\% & 0.061 & 0.094 & 0.112 & 0.010 & 0.081 & 0.105 & 0.090 & 0.093 & 0.103  \\
+2\% & 0.122 & 0.158 & 0.182 & 0.088 & 0.156 & 0.205 & 0.170 & 0.192 & 0.190  \\
\hline
A(Al) from Al I lines\\
-2\% & -0.110 & -0.127 & -0.117 & -0.143 & -0.157 & -0.126 & -0.223 & -0.156 & -0.177   \\
-1\% & -0.058 & -0.062 & -0.071 & -0.089 & -0.076 & -0.042 & -0.056 & -0.092 & -0.083  \\
+1\% & 0.057 & 0.054 & 0.059 & 0.061 & 0.071 & 0.048 & 0.048 & 0.021 & 0.042   \\
+2\% & 0.106 & 0.112 & 0.121 & 0.129 & 0.141 & 0.087 & 0.090 & 0.068 & 0.100  \\
\hline
A(Si) from Si I lines\\
-2\% & -0.140 & -0.103 & -0.071 & -0.177 & -0.226 & -0.284 & -0.447 & -0.224 & -0.224  \\
-1\% & -0.071 & -0.080 & -0.032 & -0.074 & -0.115 & -0.146 & -0.211 & -0.099 & -0.099  \\
+1\% & 0.072 & 0.084 & 0.092 & 0.124 & 0.106 & 0.133 & 0.192 & 0.093 & 0.093  \\
+2\% & 0.131 & 0.118 & 0.162 & 0.211 & 0.196 & 0.240 & 0.325 & 0.192 & 0.192  \\
\hline
A(K) from K I lines\\
-2\% & -0.101 & -0.105 & -0.111 & -0.118 & -0.103 & -0.103 & -0.099 & -0.098 & -0.093  \\
-1\% & -0.050 & -0.051 & -0.055 & -0.057 & -0.050 & -0.050 & -0.048 & -0.048 & -0.046  \\
+1\% & 0.047 & 0.049 & 0.052 & 0.055 & 0.048 & 0.048 & 0.047 & 0.046 & 0.045  \\
+2\% & 0.091 & 0.096 & 0.102 & 0.107 & 0.095 & 0.095 & 0.092 & 0.091 & 0.088  \\
\hline
A(Ca) from Ca I lines\\
-2\% & -0.070 & -0.083 & -0.082 & -0.096 & -0.080 & -0.099 & -0.123 & -0.139 & -0.106   \\
-1\% & -0.040 & -0.040 & -0.039 & -0.047 & -0.043 & -0.053 & -0.066 & -0.081 & -0.100  \\
+1\% & 0.029 & 0.040 & 0.040 & 0.046 & 0.039 & 0.050 & 0.052 & 0.061 & 0.093  \\
+2\% & 0.068 & 0.082 & 0.082 & 0.094 & 0.082 & 0.103 & 0.111 & 0.136 & 0.201  \\
\hline
A(Ti) from Ti I lines\\
-2\% & -0.147 & -0.150 & -0.152 & -0.166 & -0.137 & -0.126 & -0.129 & -0.134 & -0.138   \\
-1\% & -0.073 & -0.074 & -0.075 & -0.082 & -0.070 & -0.063 & -0.065 & -0.069 & -0.070  \\
+1\% & 0.060 & 0.063 & 0.063 & 0.070 & 0.067 & 0.062 & 0.066 & 0.068 & 0.070   \\
+2\% & 0.115 & 0.119 & 0.121 & 0.132 & 0.128 & 0.119 & 0.127 & 0.132 & 0.136   \\
\hline
A(V) from V I lines\\
-2\% & -0.265 & -0.265 & -0.281 & -0.338 & -0.365 & -0.336 &  & &   \\
-1\% & -0.124 & -0.125 & -0.135 & -0.192 & -0.209 & -0.185 & & &   \\
+1\% & 0.084 & 0.089 & 0.099 & 0.164 & 0.157 & 0.232 & & &   \\
+2\% & 0.139 & 0.147 & 0.162 & 0.253 & 0.237 & 0.394 & & &   \\
\hline
A(Cr) from Cr I lines\\
-2\% & -0.193 & -0.219 & -0.267 & -0.310 &  & & & &   \\
-1\% & -0.091 & -0.103 & -0.126 & -0.208 & & & & &   \\
+1\% & 0.074 & 0.083 & 0.099 & 0.169 & & & & &   \\
+2\% & 0.130 & 0.144 & 0.168 & 0.261 & & & & &   \\
\hline
A(Mn) from Mn I lines\\
-2\% & -0.146 & -0.172 & -0.212 & -0.306 & -0.427 & -0.541 & & &   \\
-1\% & -0.071 & -0.083 & -0.101 & -0.143 & -0.190 & -0.265 & &   \\
+1\% & 0.064 & 0.072 & 0.085 & 0.104 & 0.122 & 0.140 & & &   \\
+2\% & 0.121 & 0.137 & 0.156 & 0.180 & 0.208 & 0.245 & & &   \\
\hline
A(Ni) from Ni I lines\\
-2\% & -0.500 & -0.456 & -0.490 & -0.574 & & & & &   \\
-1\% & -0.345 & -0.272 & -0.304 & -0.342 & & & & &   \\
+1\% & 0.223 & 0.277 & 0.264 & 0.248 & & & & &   \\
+2\% & 0.321 & 0.398 & 0.411 & 0.327 & & & & &   \\
\hline
\tablewidth{0pt}	
\enddata
\tablenotetext{}{}
\end{deluxetable*}

%%%%%%%%%%%%%%%%%%%%%%%%%%%%%%%%%%%%%%%%%%%%%%%%%%

\label{lastpage}

{}

\end{document}